\documentstyle[aaspp4]{article}

\def\be{\begin{equation}}
\def\ee{\end{equation}}
\def\ba{\begin{eqnarray}}
\def\ea{\end{eqnarray}}

\journalid{VOL}{JOURNAL DATE}
\articleid{START PAGE}{END PAGE}
\paperid{MANUSCRIPT ID}

\slugcomment{To appear in the Astrophysical Journal}

\begin{document}

\title{Relativistic precession around rotating neutron stars: Effects
	due to frame-dragging and stellar oblateness}
\author{Sharon M. Morsink}
\affil{Department of Physics, University of Wisconsin-Milwaukee, 
P.O. Box 413, Milwaukee, WI 53201 \\
morsink@csd.uwm.edu }

\and

\author{Luigi Stella}
\affil{
Astronomical Observatory of Rome, Via Frascati 33 \\
00040 Monteporzio Catone (Roma) \\
stella@coma.mporzio.astro.it \\ 
Affiliated to ICRA }

\begin{abstract} 
General relativity predicts that a rotating body produces a frame-dragging
(or Lense-Thirring) effect: the orbital plane of a test particle in a
non-equatorial orbit precesses about the body's symmetry axis. In this paper
we compute the precession frequencies of circular orbits around rapidly
rotating neutron stars for a variety of masses and equations of state. 
The precession frequencies computed are expressed
as numerical functions of the orbital frequency observed at infinity. The 
post-Newtonian expansion of the exact precession formula is examined
to identify the relative magnitudes of the precession caused by
the Lense-Thirring effect, the usual Newtonian quadrupole effect and 
relativistic corrections. The first post-Newtonian correction to the
Newtonian quadrupole precession is derived in the limit of slow rotation. 
We show that the post-Newtonian precession formula is a good approximation 
to the exact precession close to the neutron star in the slow rotation limit
(up to $\sim 400$~Hz in the present context). 

The results are applied to recent RXTE observations of neutron star 
low-mass X-ray binaries, which display kHz quasi-periodic oscillations 
and, within the framework of beat frequency models, allow the measurement
of both 
the neutron star spin frequency and the Keplerian frequency of the innermost 
ring of matter in the accretion disk around it. 
For a wide range of realistic equations of state, we find that the predicted
precession frequency of this ring is close to one half of the 
low-frequency ($\sim 20 - 35$ Hz) quasi-periodic oscillations 
seen in several Atoll sources.
\end{abstract}


%%%%%%%%%%%%%%%%%%%%%%%%%%%%%%%%%%%%%%%%%%%%%%%%%%%%%%%%%%%%%%%%%%%%%%
%%%%%%%%
\section{Introduction}
\label{s:introduction}

Neutron stars and black holes are the most compact objects, as such they 
provide an arena for the observation of strong gravitational effects
predicted by the general theory of relativity. 
An important relativistic effect is the dragging of inertial frames,
first predicted by Lense and Thirring (Lense and Thirring 1917). The
Lense-Thirring frame dragging causes zero angular momentum observers
(ZAMO's) to orbit around a  rotating body with non-zero angular velocity. If
the orbit is inclined to the  body's equatorial plane, the plane of the
orbit will precess. It has recently been noted (Stella and Vietri 1998a) that
the precession frequency due to the  frame-dragging effect has  
approximately the right magnitude and radial dependence 
to explain observations of accreting neutron stars made by the Rossi
X-ray Timing Explorer (RXTE) satellite. If  this interpretation is 
confirmed, it would provide a confirmation of the Lense-Thirring effect 
that includes its radial dependence, a test that is beyond 
reach of any other current or planned experiment to test 
frame dragging. 
The other interesting implication of this  proposal is that it may be
possible to constrain the neutron star's equation of state, since the
magnitude of the effect depends strongly on the star's  density profile. In
order to determine whether the frame-dragging effect has been observed, the
details of the dependence of the precession rate on  position and the star's
mass, angular velocity and equation of state must be known. In this paper we
will numerically compute the precession frequencies predicted by general
relativity. 

The magnitude of the Lense-Thirring effect increases with the rotating body's 
angular momentum and compactness, so that it is largest for rapidly rotating
neutron stars and black holes. 
%LS begin
However, the 
measurement of the frame-dragging due to the Earth's motion 
appears to be within reach and several groups are actively pursuing this 
possibility.  
The Gravity Probe B experiment (Everitt et al. 1993 and references therein)
is designed to measure  the precession of gyroscopes in orbit around the Earth
to an accuracy in the percent range, 
while a lower accuracy measurement of the Lense-Thirring precession of 
the orbit of the LAGEOS satellites 
has been reported (Ciufolini et al. 1998 and references therein). 
%LS end
Both experiments are very difficult since the frame-dragging 
precession is many orders of magnitude smaller than the precession due to the
Earth's oblateness predicted by Newtonian gravity. The possibility of measuring
frame-dragging due to the motion of neutron stars or black holes is attractive
since the precession due to oblateness, while still important is smaller than 
the frame-dragging precession. 

RXTE has recently discovered kilohertz quasi-periodic oscillations 
(kHz QPOs) in
the X-ray light curves of 
over 15 accreting neutron stars in low mass X-ray binaries 
(see van der Klis 1998 for a review). This discovery is remarkable, since
the dynamical frequency of a particle in circular 
orbit at radius $r \sim 6 - 8 M$ 
(in geometric units G=c=1) is in the range of $1 - 1.5$ kHz for a $1.4 
M_\odot$ 
non-rotating neutron star. Thus, it is now possible to observe the region 
close to a neutron star where the gravitational field is strong and effects 
due to general relativity are important. 

The X-ray observations have shown that the kHz QPO power spectrum peaks 
typically occur in pairs. 
The frequency of 
each QPO peak varies with time, 
often in response to X-ray flux, and therefore mass accretion rate, 
variations. 
However, in nearly all cases, 
the separation between the QPO peaks remains constant.
A number of these neutron stars belong to the ``Atoll" class and emit 
sporadic Type I X-ray bursts, likely originating from thermonuclear flashes 
in the freshly accreted material on their surface.  
RXTE revealed that during X-ray bursts an additional modulation 
of the X-ray flux is present, the frequency of which is consistent 
with either one or two times the kHz QPO peak separation frequency. 
These phenomena are most readily interpreted on the basis of the 
beat-frequency modulated accretion scenario originally suggested 
to interpret the lower frequency quasi periodic signals 
observed from accreting neutron stars and white dwarfs (see e.g. van der 
Klis 1995 and references therein).
In beat frequency models the higher frequency kHz QPO peak
corresponds to the orbital frequency of clumps of matter 
at the inner boundary of the accretion disk. The
lower frequency kHz QPO peak is the beat frequency, corresponding to the 
difference between the clumps' orbital frequency 
and the star's spin frequency. This model 
naturally leads to the conclusion that the peak separation should be constant
and equal to the star's spin frequency. Application of the beat frequency
model to the accreting neutron stars observed by RXTE indicates that 
they are spinning with periods close to $3 ms$, 
although the modulation observed during some 
of type I X-ray bursts suggests that they might be spinning twice as fast.

Beat frequency models differ mainly with respect to the physical mechanism 
that determines the inner boundary of the accretion disk. This might be 
the presence of a neutron star magnetosphere (Alpar \& Shaham 1985;
Lamb et al. 1985), or the supersonic radial velocity of the disk material 
induced by radiation drag or the motion inside the 
innermost stable circular orbit, ISCO 
(Miller, Lamb \& Psaltis 1998).

Another important aspect of the phenomenology of kHz QPO sources is that 
they display broad lower frequency QPO peaks, the frequency of which is 
correlated with the frequency of the kHz QPO peaks. 
In three ``Atoll" sources $\sim 20-35$~Hz QPOs are observed; in the 
case of 4U1728-34 the frequency of these QPOs varies roughly with the 
square of the higher kHz QPO frequency (Stella and Vietri 1998a). 
In the higher luminosity ``Z"-type sources GX5-1 and GX17+2 the frequency of 
the $\sim 15-60$~Hz ``horizontal branch" QPOs appears also to depend
quadratically on the frequency of the higher kHz QPOs
(Stella and Vietri 1998a,b).
It is this correlation between peaks which has suggested
the following modification of the beat frequency
model: if the star's accretion disk is tilted with respect to the star's 
equatorial plane, the orbital plane will precess due to the combined effect
of frame-dragging and stellar oblateness. For orbits  near a neutron star
spinning with a period of $3 ms$, the frame dragging effect is dominant and 
yields precession frequencies of a few tens of Hz which vary approximately 
as the square of the orbital frequency (Stella and Vietri 1998a). Close to the
star the precession due to oblateness and other corrections due to general
relativity will also be important and will cause departures from the quadratic
scaling law. The principal purpose of this paper is to investigate in greater 
detail 
%LS 
the precession frequency and its dependence 
on orbital frequency predicted by general
relativity. This will provide a more accurate means of testing the precession
hypothesis using many simultaneous observations of the QPO peaks in both the 
tens of Hz and kHz range. 

QPO peaks have also been observed in X-ray binaries which probably contain 
black holes. 
It has been proposed that some of these QPO peaks also be identified
with disk precession (Cui et al. 1998 and references therein). 
Rotating black holes are much 
simpler than neutron stars since their gravitational fields are completely
specified by the Kerr geometry depending on only the hole's mass and angular
momentum. For this reason, measurements of precession near black holes have
the potential to provide a much cleaner measurement of the Lense-Thirring
effect. 
%LS begin
However, it should be noted that additional (kHz) QPO peaks corresponding to 
the $\phi$ and $\theta$ periodicities of spherical orbits 
have not yet been observed yet 
in black hole candidates, so that the 
precession hypothesis rests upon scarcer evidence and 
quantitative testing based on multiple QPO peaks is currently 
not a possibility.
%LS end 
In any case, a thorough investigation of the precession of 
spherical orbits in the Kerr black hole geometry has been performed 
(Wilkins 1972, Merloni et al. 1998), so  
we will instead concentrate here on neutron star spacetimes. 

In the calculations presented in this paper, it will be assumed that 
general relativity is the correct theory of gravity. The frame-dragging
precession could hypothetically be used to test the validity of general 
relativity or other alternative theories of gravity using the parameterized
post-Newtonian framework (Will 1993). However, the uncertainty in the 
equation of state for neutron stars would render it very difficult (or 
impossible) to differentiate between alternate theories of gravity.

We restrict our attention to the simplest possible model, a thin Keplerian
disk  tilted infinitesimally from the star's 
equatorial plane. 
Both the corotating and counterrotating cases will be 
presented and discussed. 
Although this model may be oversimplified, it should suffice
to capture the essential physical features of the problem and place 
accurate limits on the precession frequencies predicted by general relativity.

Accurate codes (Komatsu et. al. 1989, Cook et. al. 1992, Salgado et. al.
1994) now exist which can integrate the Einstein field equations for 
rapidly rotating neutron stars given a barotropic perfect fluid equation 
of state. Our method is to compute the spacetime metric of the neutron star
using a code written by Stergioulas (1995) which is 
equivalent to Cook et. al.'s code  and accurate to at least $1\%$ 
(Stergioulas and Friedman 1995, Nozawa et. al. 1998).  
The geodesic equations for test particles in the numerical spacetime can 
then be solved, yielding the 
orbital and precession frequencies. 

The paper is organized as follows. In section ~\ref{s:models}
 the predictions of general
relativity regarding the orbital motion of a precessing test particle are
reviewed and the numerical methods used to solve the characteristic 
frequencies are outlined. In section ~\ref{s:asymptotic} an asymptotic 
expansion of the precession formula is derived in order to discuss the 
validity of the approximate Newtonian formula. 
The neutron star parameters which were selected are described in 
section ~\ref{s:description}.  In section ~\ref{s:general}  general results
are presented for a variety of equations of state, stellar masses and 
angular velocities. These results are compared with the power spectra of the
LMXBs observed by RXTE.  We find that for corotating orbits 
the observed precession frequencies are somewhat lower than predicted 
using the approximate method of Stella and Vietri (1998a), 
therefore increasing the 
discrepancy between the predicted precession frequency and the observations,  
especially for ``Z"-type kHz QPO sources. 
These conclusions and their consequences are discussed in the final 
section. 

\section{Models of Rapidly Rotating Relativistic Neutron Stars}
\label{s:models}
The neutron star models which we compute are assumed to be stationary, 
axisymmetric, uniformly rotating perfect fluid solutions of the Einstein
field equations. The assumptions of stationarity and axisymmetry allow the
introduction of two commuting Killing vectors $\phi^\alpha$ and $t^\alpha$, 
generating rotations and time-translations as well as coordinates $\phi$
and $ t$, defined by the conditions (see eg., Friedman, Ipser and Parker, 1986)
\be
t^\alpha \nabla_\alpha t = \phi^\alpha \nabla_\alpha \phi = 1 \;, \;
t^\alpha \nabla_\alpha \phi = \phi^\alpha \nabla_\alpha t = 0,
\ee
labeling the two-surfaces spanned by the Killing vectors. As a result, 
the metric, $g_{\alpha \beta}$ can be written
as 
\begin{equation}
ds^2 = - e^{\gamma + \rho} dt^2 
	+ e^{\gamma - \rho} \bar{r}^2 \sin ^2 \theta \left(
		d\phi - \omega dt \right)^2 
	+ e^{2 \alpha} \left( d\bar{r}^2 + \bar{r}^2 d\theta^2\right),
\end{equation}
where the metric potentials $\rho, \gamma, \alpha$ and $\omega$ depend only
on the coordinates $\bar{r}$ and $\theta$. 
The function $\frac{1}{2}(\gamma + \rho)$
 is the relativistic
generalization of the Newtonian gravitational potential; the time dilation
factor between an observer moving with angular velocity $\omega$ and an
observer at infinity is $e^{\frac{1}{2}(\gamma + \rho)}$.
 The coordinate $\bar{r}$ is not the same as the Schwarzschild
coordinate $r$. In the limit of spherical symmetry, 
$\bar{r}$ corresponds to the
isotropic Schwarzschild coordinate. Circles centred about the axis of 
symmetry have circumference $2 \pi r$ where $r$ is related to our 
coordinates $\bar{r},\theta$ by 
\be
r = e^{\frac{1}{2}(\gamma - \rho)} \bar{r} \sin \theta .
\label{radius}
\ee
The metric potential $\omega$ is the angular velocity about the symmetry
axis  of ZAMOs and is responsible for the Lense-Thirring effect. 
The fourth metric potential, $\alpha$ specifies the geometry of the 
two-surfaces of constant $t$ and $\phi$. When the star is non-rotating, the
exterior geometry is that of the isotropic Schwarzschild metric, with
\be
e^{\frac{1}{2}(\gamma + \rho)} = \frac{1-M/2\bar{r}}{1+M/2\bar{r}}, \;
e^{\frac{1}{2}(\gamma - \rho)} = e^\alpha = \left( 1 + M/2\bar{r} \right)^2,
 \; \omega=0.
\label{isotropic}
\ee

We shall investigate uniformly rotating perfect fluid stars with stress
tensor
\be
T^{\alpha \beta} = (\epsilon + p) u^\alpha u^\beta 
	+ p g^{\alpha \beta}
\ee
where $\epsilon$ and $p$ are the fluid's energy density and pressure as 
measured by a co-moving observer. The fluid four-velocity, 
\be
u^\alpha = u^t ( t^\alpha + \Omega \phi^\alpha), 
\ee
is a linear combination of the Killing vectors. The  factor $u^t$ is a 
function of $r$ and $\theta$ specified by the condition that $u^\alpha$ is
a unit, time-like vector, while $\Omega$ is the constant angular velocity 
of the star. We will numerically integrate the field equations by assuming
various zero-temperature, barotropic equations of state of the form
$\epsilon = \epsilon(p)$. 

The numerical method used to integrate the Einstein field equations for the
metric potentials is identical to the method described by Cook et. al. 
(1992 and corrections in 1994a) which is based on the self-consistent
field method presented by 
Komatsu et. al. (1989). Since the method has been described in detail in 
these papers, we will only explicitly discuss aspects of the code which have
direct bearing on the present work. 

The field equations can be written in a form of a set of elliptic partial
differential equations, which can be formally solved using a standard Green 
function approach (Komatsu et. al. 1989). As a result, the solution for
the metric potentials $\nu, \beta$ and $\omega$ can be expressed as a series
\ba
\gamma(\bar{r},\theta) &=&
	- \frac{2}{\pi} e^{-\gamma/2} \Sigma_{n=1}^{\infty} 
	\frac{ \sin \left( (2n-1) \theta \right)} 
	{(2n-1) \sin \theta} \gamma_{n+1}(\bar{r})  \label{gamma}   \\
\rho(\bar{r},\theta) &=&
	-  e^{-\gamma/2} \Sigma_{n=0}^{\infty}
	P_{2n}( \cos \theta) \rho_{n+1}(\bar{r})     \label{rho} \\
\omega(\bar{r},\theta) &=& - e^{\rho - \gamma/2}   \Sigma_{n=1}^{\infty} 
	\frac{ P^1_{2n-1}(\cos \theta)}
	{2n(2n-1) \sin \theta} \omega_{n+1}(\bar{r}) \label{omega}
\ea
where the expansion coefficients, given by integrals of the metric potentials
over all of space are determined iteratively once an equation of state is
specified. In the numerical implementation,
the sums are taken only over a finite range and we keep only the first 
ten terms  in the series. The fourth metric potential, 
$\alpha$ is determined by 
quadrature once the other potentials are known. 

Once the geometry of the star is determined, the total mass, baryonic mass 
 and angular momentum
of the star, $M$, $M_B$ and $J$ respectively can be computed. This involves 
integration of the metric potentials and the stress tensor over all space 
(Komatsu et. al. 1989).

\subsection{Geodesic Motion for Circular Orbits}

The assumptions that the neutron star is stationary and axisymmetric has the
consequence that two constants of motion are associated with any orbit, 
corresponding to the energy per unit mass, $E$, and orbital angular momentum
per unit mass, $L$. As a result, the set of all circular orbits confined
to the star's equatorial plane is a two-parameter family depending only 
on $E$ and $L$, once the background metric is determined. The theory of 
these orbits and their perturbations is presented by Bardeen (1970, 1973). 
Bardeen introduces a one-parameter family of orbits corresponding to the 
motion of zero-angular momentum observers (ZAMOs) which have $L=0$. Although
these observers have zero angular momentum, the dragging of inertial frames
by the rotating star causes the ZAMOs to orbit the star with angular 
velocity $\omega$, as seen by an observer at infinity. 

It is useful to consider the motion of other observers with reference to 
the ZAMOs. Consider a particle confined to a circular orbit in the
star's equatorial plane with four-velocity
\be
U^\alpha := \frac{dx^\alpha}{d\tau} = \left[\frac{dt}{d\tau}\right]_{\pi/2}
	 \left(
	t^\alpha + \left[\frac{v_\infty}{r}\right]_{\pi/2}
	 \phi^\alpha \right)
\ee
where $\tau$ is the particle's proper time and $v_\infty$ is the particle's
physical three-velocity as measured by an observer at infinity. Square 
brackets with subscript $\pi/2$ denotes that the quantity within the brackets
is evaluated in the star's equatorial plane, $\theta = \pi/2$. 
The  four-momentum per unit particle mass is 
\be
p_\alpha = g_{\alpha \beta} U^\beta = - E \nabla_\alpha t 
	+ L \nabla_\alpha \phi \;.
\ee
>From the definition of four-momentum, the particle's four-velocity is 
determined by $E$ and $L$ through the relations
\ba
\left[\frac{dt}{d\tau}\right]_{\pi/2}  &=& \left[
	(E-\omega L) e^{-(\gamma+\rho)}\right]_{\pi/2} \label{redshift}\\
\left[ v_\infty\right]_{\pi/2}  &=& \left[\omega r + \frac{L}{E-\omega L} 
	\frac{e^{\gamma + \rho}}{r} \right]_{\pi/2}
\label{vinfty}
\ea

A ZAMO will
measure the particle to move with a velocity, $v$ defined by
\be 
\left[ v\right]_{\pi/2} = 
	\left[e^{-(\gamma + \rho)/2}( v_\infty - \omega r )\right]_{\pi/2}.
\label{vinfinity}
\ee
The  constants of motion $E$ and $L$ can  be rewritten explicitly in terms
 of $v$, 
\ba
L &=&\left[ \frac{ v r}{\sqrt{1-v^2}}\right]_{\pi/2} \\
E -\left[ \omega\right]_{\pi/2} L  &=& 
	\left[ \frac{ 1}{\sqrt{1-v^2}} 
	e^{\frac{1}{2}(\gamma + \rho)}\right]_{\pi/2},
\ea
where $(E-\omega L) \exp{-\frac{1}{2}(\gamma + \rho)}$ is the energy of the 
particle measured by a ZAMO. 
The solution of the geodesic equation yields two possible values for the
three-velocity (Bardeen, 1970), 
\be
\left[v_{\pm}\right]_{\pi/2} = 
	\left[\frac{ e^{-\rho}\bar{r}^2\omega_{,\bar{r}} \pm 
	\left( e^{-2\rho}\bar{r}^4\omega_{,\bar{r}}^2 
	+ 2\bar{r}(\gamma_{,\bar{r}} + \rho_{,\bar{r}})
		+ \bar{r}^2(\gamma_{,\bar{r}}^2 - \rho_{,\bar{r}}^2)
	\right)^{1/2} }
	{ 2 + \bar{r} (\gamma_{,\bar{r}} - \rho_{,\bar{r}})}\right]_{\pi/2},
\label{vzamo}
\ee
where $v_+$ and $v_-$ correspond to co-rotating and counter-rotating
orbits respectively. The Kepler frequencies, $\nu_{K\pm}$ are defined as the 
orbital frequency of the prograde and retrograde 
circular orbits as measured by asymptotic observers
\be
\nu_{K\pm} :=\left[  \frac{v_{\infty\pm}}{2\pi r}\right]_{\pi/2} 
	= \left[    \frac{1}{2\pi}
	\left( v_{\pm} \frac{e^\rho}{\bar{r}} + \omega \right)\right]_{\pi/2}
\label{kepler}
\ee

\subsection{Precession of Tilted Orbits}

We now turn to circular geodesics which are not confined to the star's 
equatorial plane. It is simplest to consider motion where the angle of 
inclination between
the orbital and equatorial planes is infinitesimally small.

The four-velocity of a particle moving on a circular orbit inclined to the
star's equatorial plane is
\be
U^\alpha =   \frac{dt}{d\tau} \left(
	t^\alpha + \frac{v_\infty}{r} \phi^\alpha \right)
	+ \frac{d \theta} {d\tau} \delta^\alpha_\theta \;.
\ee
The values of $\frac{dt}{d\tau}$ and $v_\infty$ on the star's equatorial plane
are given by Eqs. (\ref{redshift}) and (\ref{vinfty}) respectively. 
 The normalization condition $U^\alpha 
U_\alpha = -1 $ leads to an equation for $\theta(\tau)$ which takes the form
of  particle motion in a one-dimensional potential,
\be
0= \frac{1}{2} \left(  \frac{d \theta} {d\tau} \right)^2 +
	V(\theta) \;,
\ee
where the potential is (Bardeen, 1970)
\be
V(\theta) = \frac{1}{2} \frac{ e^{-2\alpha}}{\bar{r}^2} 
	\left( -1 + e^{- (\gamma + \rho)}(E-\omega L)^2 
		+ \frac{L^2}{\bar{r}^2 \sin ^2 \theta} e^{\rho - \gamma} 
	\right) \;.
\ee
Consider an orbit which only makes small oscillations out of the star's 
equatorial plane. This orbit is defined by the conditions 
$\left[ V(\theta)\right]_{\pi/2}
= \left[\partial_\theta V \right]_{\pi/2} = 0$. 
The first condition is automatically 
met by the assumed form of the four-velocity. The symmetry of the spacetime
guarantees that the second condition is met. The frequency of the oscillation
is given by 
\be
\left( d\theta/ d\tau \right)^2 = \left[ \partial_\theta 
\partial_\theta
V\right]_{\pi/2} \;.
\ee  
The precession frequency $\nu_p$  of disk's orbital plane about the star's 
axis of symmetry is the difference between the frequency of oscillations
of the particle in the longitudinal and latitudinal directions, 
$2 \pi \nu_p := d\phi/dt - d\theta/dt$, as observed by observers at 
infinity. The explicit formula is
\be
2 \pi \nu_{p\pm} = \left[ \omega 
	+ \frac{v_{\pm}}{\bar{r}} e^\rho \left( 1 - 
	e^{\alpha + \frac{1}{2}(\gamma - \rho)} X(r,\theta)
	 \right) \right]_{\pi/2}
\label{precession}
\ee
where $\nu_{p+}$ and $\nu_{p-}$ are the precession frequencies of 
prograde and retrograde orbits respectively, 
and the function $X(r,\theta)$ is defined by
\be
X(r,\theta) := 
\left(1 + \frac{1}{2} \partial_\theta \partial_\theta (\rho-\gamma)
	 + \frac{1}{2 v_{\pm}^2} \partial_\theta 
	\partial_\theta(\rho+\gamma)
	+ \frac{\bar{r}}{v_\pm} e^{-\rho} 
	\partial_\theta \partial_\theta \omega
	\right)^{1/2} .
\label{x}
\ee
This formula is similar to the expression for the precession frequency 
derived by Ryan (1995) for orbits in a vacuum spacetime. Since Ryan 
assumes the spacetime is vacuum everywhere, his expression for the precession
frequency does not include terms proportional to 
$\partial_\theta \partial_\theta \gamma$.

For slowly 
rotating stars with dimensionless rotation parameter $j=J/M^2 \ll 1$,
it is meaningful to split Eq. (\ref{precession}) into two 
terms, a precession frequency due to the Lense-Thirring effect, 
\be
\nu_{LT} = (\nu_{p+} + \nu_{p-})/2 
=\omega/2\pi + O(j^3)
\label{lt_freq}
\ee
and a precession due to the star's oblateness, 
\be
\nu_{oblate} = (\nu_{p+} - \nu_{p-})/2
\label{oblate_freq}
\ee
which is the relativistic 
generalization of the 
Newtonian quadrupole precession. Thus, to leading order, the 
precession of orbits due to the L-T effect is equal to the orbital frequency
of a ZAMO ($\omega/2\pi$) in an equatorial orbit around the star. 
It should be remembered, however, that 
the frame-dragging potential $\omega$ makes contributions to the velocity,
$v_{\pm}$, and the frequency
which we have named $\nu_{oblate}$. If the star is rotating
rapidly, such a split may not be meaningful. 

The precession frequency is easily computed once the metric expansion 
coefficients appearing in Eqs. (\ref{gamma}) - (\ref{omega}) have been
computed. The second partial derivative of the metric potentials are
evaluated using standard relations for Legendre polynomials and 
trigonometric functions
\ba
\left[ \partial_\theta \partial_\theta \gamma \right]_{\pi/2} &=&
	- \left[ \frac{4}{\pi(2+\gamma)} e^{-\gamma/2}\right]_{\pi/2}
	\Sigma_{n=2}^{\infty} (-1)^{n} \left(
	(2n-1) - \frac{1}{2n-1} \right) \gamma_{n+1}(r) \\
\left[ \partial_\theta \partial_\theta \rho \right]_{\pi/2} &=&
	- \left[ \frac{\rho \partial_\theta \partial_\theta \gamma}{2}
	\right]_{\pi/2}
	- \left[  e^{-\gamma/2}\right]_{\pi/2} \Sigma_{n=1}^{\infty} 
	(2n-1)(2n+1) P_{2n-2}(0) \rho_{n+1}(r) \\
\left[ \partial_\theta \partial_\theta \omega \right]_{\pi/2} &=&
	\left[ \omega \left(\partial_\theta \partial_\theta \rho
		- \frac{ \partial_\theta \partial_\theta \gamma}{2} \right)
	\right]_{\pi/2} \nonumber \\
	&&
	- \left[  e^{\rho - \frac{1}{2}\gamma} \right]_{\pi/2}
	\Sigma_{n=2}^{\infty} \left( \frac{1}{n} - (2n-1) \right)
	P_{2n-2}(0) \omega_{n+1}(r) \;.
\ea

\section{Asymptotic Expansions}
\label{s:asymptotic}

Our main interest in this paper is in the motion of test particles near a 
neutron star rotating arbitrarily quickly. The motion of a particle making 
small oscillations out of the star's equatorial plane is determined 
exactly by the formulae presented in section \ref{s:models}, once the 
spacetime's geometry has been determined numerically. Although, this method
provides a sufficient  description of the motion, it is useful to examine
the weak field, slow rotation limit to elucidate the nature of the orbital
precession. This corresponds to a double expansion in two small parameters,
$M/r$ and $j:=J/M^2$, which isn't generally a good description of the 
strong gravitational field near a rapidly rotating neutron star. However, 
from this double expansion, the Newtonian quadrupole formula will emerge,
which will allow us to discuss the validity of the 
Newtonian formula.

Since the spacetime is asymptotically flat, the coefficients in the expansion 
of the metric potentials Eqs. (\ref{gamma}) - (\ref{omega}) 
must decay  at large $r$ as (Komatsu et. al. 1989)
\be
\gamma_{n+1}(r) \sim \left(\frac{M}{r}\right)^{2n} \;,\;
\rho_{n+1}(r) \sim \omega_{n+1}(r) \sim  \left(\frac{M}{r}\right)^{2n+1}\;.
\ee
In the slow rotation approximation is an expansion in small $j$ with
leading order behaviour of the coefficients given by 
\be
\gamma_{n+1}(r) \sim \rho_{n+1}(r) \sim j^{2n} \;, \; 
	\omega_{n+1}(r) \sim j^{2n-1} \;.
\ee
Hartle's (1967) slow rotation formalism is equivalent to keeping terms in 
the series expansions (\ref{gamma}) - (\ref{omega}) up to and including 
$n=1$. The explicit form of the leading order coefficients have been calculated
by Butterworth and Ipser (1976). Comparing with their work, we find
the following dependencies at large $r$, keeping only the terms of lowest
order in $j$,
\ba
\rho_1(\bar{r}) &=& \frac{2M}{\bar{r}} + \frac{B_0}{\bar{r}^2}\left(1 + \frac{M}{3\bar{r}}\right) +
	O(1/\bar{r}^4) \label{rho1} \\
\rho_2(\bar{r}) &=& - 2 \frac{\Phi_2}{\bar{r}^3} + 4 \frac{J^2}{\bar{r}^4}+ O(1/\bar{r}^5)
	\label{rho2} \\
\gamma_2(\bar{r}) &=& - \frac{\pi}{2} \frac{B_0}{\bar{r}^2} + O(1/\bar{r}^3)\label{gamma2} \\
\omega_2(\bar{r}) &=& - 4\frac{J}{\bar{r}^3} + O(1/\bar{r}^4) \label{omega2}
\ea
where $\Phi_2$ is the relativistic generalization of the Newtonian 
quadrupole moment (Butterworth and Ipser 1976; Ryan 1995; 
Laarakkers and Poisson 1997). We choose to define $\Phi_2$ so that it is 
positive for an oblate spheroid. The second term in Eq. (\ref{rho2}) is 
the first post-Newtonian correction to the quadrupole moment. The 
constant $B_0$ is determined by an integral over the star given by Butterworth
and Ipser (1976) (which they label $\tilde{B}_0$). Integration of the 
field equation for $\alpha$ at large $\bar{r}$ results in
\be
\alpha + \frac{1}{2}(\gamma + \rho) = \frac{1}{\bar{r}^2} 
	\left( - B_0 - \frac{M^2}{2} + \cos^2 \theta \left( 2 B_0 + 
		\frac{M^2}{2} \right) \right)
	+ O(1/\bar{r}^4) \;.
\label{alpha}
\ee 

 Substituting the asymptotic expansions (\ref{rho1}) - (\ref{omega2})
into Eq. (\ref{vzamo}), the particle velocity measured by observers at  
infinity is 
\be
v_{\infty \pm} = \pm \sqrt{ \frac{M}{r}} - j \frac{M^2}{r^2}
	+ O\left( j^2 
	 \left( \frac{M}{r}\right)^{5/2} \right) .
\label{vasymptotic}
\ee
The number of terms which must be
included in this  series depends on the relative size of the two expansion 
parameters. For $1.4 M_\odot$ stars spinning at the rate observed by RXTE, 
( $\sim$ 3 ms period), the rotation parameter $ j \sim 0.1 - 0.2$ for 
equations of state
from the Arnett and Bowers (1977) catalog (Cook et. al. 1994). 
For this range of rotation
parameter, when  $M/r < 4/9$ the second term in Eq. (\ref{vasymptotic}) is 
at most a $6\%$ correction to the first term, which is the usual Kepler
velocity for a non-rotating spacetime. However, if the star is rotating 
close to its maximum possible angular velocity, then the rotation parameter
may approach unity (in fact there is no theoretical upper limit for $j$
as there is for black holes) and Eq. (\ref{vasymptotic}) will not be a good 
approximation for Eq. (\ref{vinfinity}).

A weak field, slow rotation limit of the precession frequencies  can be 
derived in the limit $j \left( M/r \right)^{3/2} \ll 1$. In this case, 
occurences of $v_{\pm}$ and $ \bar{r}$ can be replaced by $\nu_{K}$ and the 
star's mass, through the relation
\be
\frac{v e^\rho}{\bar{r}} = 2 \pi \nu_K 
	+ O( j \left( \frac{M}{r} \right)^{3/2} )
 	= \sqrt{ \frac{M}{r^3}} 
	+ O( j \left( \frac{M}{r} \right)^{3/2} ) .
\label{kepler2}
\ee
In this limit, the Lense-Thirring precession frequency, Eq. (\ref{lt_freq})
is 
\be
\nu_{LT} = \frac{8 \pi^2  I \nu_s \nu_K^2}{c^2 M} ,
\label{LT} 
\ee
where factors of $G$ and $c$ have been restored and 
$\nu_s := \Omega/2\pi$ is the spin frequency of the star.

%%%%%%%%%%%%%%%%%%%%%%%%%%%%%%%%%%%%%%%%%%%%%%%%%%%
The precession due to the star's oblateness can be found by noting that 
in the weak field, slow rotation limit, 
the function $X$ appearing in Eq. (\ref{precession}) is
\be
[ X ]_{\pi/2} = \left( 1 + {3} \frac{ (v^2 + 1)}{v^2} 
	\left( \frac{ \Phi_2}{ \bar{r}^3 } - 2 \frac{ J^2}{\bar{r}^4}
	\right) \right)^{1/2} .
\ee 
Making use of the asymptotic expression Eq. (\ref{kepler2}) for $v$,
the function $X$ reduces to 
\be
[ X ]_{\pi/2} =  1 + {3} \frac{ e^{2\rho} r^3}{M \bar{r}^2}
	\left( 1 + \frac{M}{\bar{r}} \right)
	\left( \frac{ \Phi_2}{ \bar{r}^3 } - 2 \frac{ J^2}{\bar{r}^4}
	\right)  .
\ee 
The expression for $\nu_{oblate}$ is then
\ba
\nu_{oblate} &=& \nu_K 
	\left(  1 - e^{-\alpha + \frac{1}{2}(\gamma -\rho)} \right)  
	\nonumber \\
	&& - 3 \nu_K  \frac{ 
	e^{2\rho + \alpha + \frac{1}{2}(\gamma - \rho)} r^3}
	{M \bar{r}^2}
	\left( 1 + \frac{M}{\bar{r}} \right)
	\left( \frac{ \Phi_2}{ \bar{r}^3 } - 2 \frac{ J^2}{\bar{r}^4}
	\right) .
\ea
At this level of approximation it is sufficient to use the values of 
$\bar{r}, \rho, \gamma$ and $\alpha$ in spherical symmetry, given by 
Eqs. (\ref{radius}) and (\ref{isotropic}).
The precession frequency due to the star's oblateness can now 
be split into terms
\be
\nu_{oblate} =  \nu_{quad}\left( 1 + c_{PN} \frac{M}{r}\right) 
		+ \nu_{cent}  .
\label{expansion}
\ee
 In equation (\ref{expansion}),
$\nu_{quad}$ and $\nu_{cent}$ correspond to the the Newtonian quadrupole
precession and a centrifugal precession due to other relativistic effects 
which enter at the same order as the Newtonian quadrupole term. The
constant $c_{PN}$ is the coefficient of the first post-Newtonian 
correction to the Newtonian 
quadrupole precession. The Newtonian quadrupole precession is 
\ba
\nu_{quad} &=& - \frac{3 \nu_K}{ 2} \frac{\Phi_2}{M r^2} \label{quad1}\\
	 &=& - 3 \Phi_2 \left( \frac{2\pi^2}{G^2 M^5} \right)^{1/3}
	 \nu_K^{7/3}.
	\label{quad2}
\ea
The post-Newtonian correction parameter has the value
\be
c_{PN} =  3 - 2 j^2 \frac{ M^3}{ \Phi_2}  \;.
\label{cpn}
\ee
 For black holes 
$\Phi_2 = \frac{G^2}{c^4} j^2 M^3$, so that $c_{PN}=1$. 
However, for neutron stars
this is not the case. It has been shown by Laarakkers and Poisson (1997)
that $\Phi_2 \simeq a(M,\hbox{EOS}) \frac{G^2}{c^4} j^2 M^3$ 
with $a(M,\hbox{EOS})$ a constant which depends on the star's mass and 
equation of state. For EOS from the Arnett and Bowers (1977) catalogue,
$a \sim 2 - 12$. The constants $a(M,\hbox{EOS})$ increase with decreasing
mass for a fixed EOS, while they increase for increasing stiffness for 
fixed mass (Laarakkers and Poisson 1997). 
It follows that the  first post-Newtonian 
correction term is most important for the stiffest equations of state and
the smallest mass stars. 
At distances close to $6M$, the post-Newtonian correction term can 
correspond to as much as a $50\%$ correction to the quadrupole precession, 
and can't be neglected.

The centrifugal precession frequency is defined by 
\be
\nu_{cent} := \nu_K \left[
	\left( 1 - e^{-\alpha + \frac{1}{2}(\gamma -\rho)} \right) 
	\right]_{\pi/2} \;.
\ee
 The function $ \left[ 1-
\exp({-\alpha + \frac{1}{2}(\gamma -\rho)} )\right]_{\pi/2}$ measures
the fractional difference in the proper length elements of two 
infinitesimal  curves, both centred about the same point on the star's 
equatorial plane.
 The first curve is defined by   
 constant values of the coordinates $\phi, \bar{r}$ and $t$, and has 
proper length $dl_1 =  \bar{r} \left[  \exp\alpha\right]_{\pi/2} d\theta$. 
The second curve is restricted to 
the star's 
equatorial plane and has proper length 
$dl_2 =  \bar{r} \left[  \exp{\frac{1}{2}(\gamma -\rho)}\right]_{\pi/2} 
d\phi$.  
If the spacetime is spherically symmetric (or flat), the proper
length of the two curves must agree and the frequency $\nu_{cent}$ 
vanishes. When the star is rotating, the proper lengths will not agree, but
the condition of asymptotic flatness requires that their difference falls off
at least as fast as $1/r^2$. Making use of the asymptotic expansions 
(\ref{rho1}) - (\ref{alpha}), the centrifugal precession frequency is 
\be
\nu_{cent} =  - \nu_K \beta \left(\frac{M}{ r}\right)^2 
\ee
where the dimensionless constant $\beta$  defined by
\be
\beta = \frac{1}{M^2}\left(2 B_0 + M^2/2 \right)
\ee
is of order $j^2$ for slowly rotating stars. The percent 
magnitude of the centrifugal 
precession relative to the quadrupole precession is 
\be
\zeta := \frac{\nu_{cent}}{\nu_{quad}} \times 100 = 
	\frac{200}{3} \beta \frac{M^3}{\Phi_2}
\label{zeta}
\ee
and is displayed in Tables \ref{t:tableA} - \ref{t:tableL}. 
The size of this term relative
to the quadrupole precession depends on the star's mass and 
equation of state, but 
typically corresponds to a correction of $1 - 15 \%$. It should be noted 
that the split between centrifugal and quadrupole precession made here is 
{\em not} coordinate invariant.

%%%%%%%%%%%%%%%%%%%%%%%%%%%
%%%%%%%%%%%%%%%%%%%%%%%%%%%

\section{Description of neutron star models selected}
\label{s:description}
 
Any theoretical predictions involving neutron stars are by necessity highly 
uncertain, due to our lack of knowledge of the behaviour of matter at high 
densities. Compounding this uncertainty is the difficulty in making
simultaneous measurements of the star's mass, radius and angular velocity. 
In order to discuss the expected precession rates predicted by general
relativity, it is necessary to carefully choose a sample of models which 
cover a reasonable volume of the allowed parameter space. 

\subsection{Equations of state}

We have chosen a set of four equations of state: A, WFF3, C, L (in order of
increasing stiffness) which cover a range of possible neutron star properties. 
Equations of state A (Pandharipande 1971), C (Bethe and Johnson 1974) and
L (Pandharipande and Smith 1975) are labeled as in the Arnett and Bowers 
(1977) catalogue. EOS A is the softest equation of state which allows a 
stable $1.4 M_\odot$ neutron star with zero angular velocity. EOS L is the 
stiffest EOS in the Arnett and Bowers catalogue which doesn't involve 
a phase transition. EOS WFF3 is a more modern EOS computed by Wiringa et. al.
(1988) using a variational framework. They use data from nucleon-nucleon 
scattering and the known properties of nuclear matter to constrain the 
two- and three-nucleon potentials in their Hamiltonians and derive a causal
``best fit'' equation of state for high densities. The equation of state
WFF3 is a match of the EOS denoted UV14+TNI (Wiringa et. al. 1988)
 at high densities to the FPS (Lorenz et. al. 1993) 
EOS low densities and corresponds to a conservative estimate of the 
correct neutron star EOS.

\subsection{Angular velocities}

In the applications of beat frequency models to twin peak kHz QPO sources
the orbital frequency of the innermost disk region corresponds to 
the higher frequency kHz QPOs,  
while the neutron star's spin frequency $\nu_s = \Omega/2 \pi$ 
is inferred from the frequency separation of the twin peaks. The range of spin
frequencies determined from this method is approximately 
 $260 Hz \le \nu_s \le 360 Hz$. 
However, it should be noted that in 
Sco X-1 (van der Klis et al. 1997) and perhaps also 4U1608-52 (Mendez 
et al. 1998)
the peak separation significantly decreases for increasing kHz QPO frequencies.
Correspondingly beat frequency scenarios are not directly applicable to these 
neutron stars, and their spin frequency, though likely in the same range, 
is not precisely known.

In several kHz QPO sources of the ``Atoll" group an additional modulation
has been revealed during type I X-ray bursts. The frequency of this  
ranges from $\sim 360$ to $\sim 580$~Hz.  
In the ``Atoll" sources which show also the twin peak QPOs, the frequency 
of the modulation during type I bursts is consistent with either the peak 
separation or twice the peak separation, further corroborating the beat 
frequency interpretation. 
%LS begin
The peak separation frequency $355 \pm 5$ Hz and burst QPO $\sim 363$ Hz 
agree for the source 4U 1728-34 (Strohmayer et al. 1996). For the source
4U 0614+091 (Ford et. al. 1996), the peak separation of $323 \pm 4$ Hz 
agrees with the marginally detected $\sim 328$~Hz QPO peak in an interval 
of persistent emission.  
%LS end
For the source 4U 1636-53, the peak separation frequency
$276 \pm 10$ Hz (Wijnands et al. 1997) is close to half the burst frequency
of 581 Hz. Similarly, for the source KS 1731-260, the peak separation is
$260 \pm 10$ Hz (Wijnands and van der Klis 1997) while the burst frequency
is $524$ Hz (Smith et. al. 1997). 
In the sources 4U 1636-53 and KS 1731-260,
it is likely that the modulation during the bursts represents the 
second harmonic of the star's
spin frequency, but there is a possibility that these neutron stars
are spinning twice faster (i.e. at the frequency observed 
during the bursts). 
%LS begin
In 4U 1702-429, Aql X-1 and MXB 1743-29 a modulation at $\sim 330$, 
549 and 589~Hz, respectively, has been detected during type I bursts 
(Swank 1997; Zhang et al. 1998a; Strohmayer et al. 1997), but a pair of 
kHz QPO peaks has not yet been detected in the persistent emission. 

Recent RXTE observations of SAX~J1808.4-3658, a transient type I burster, 
revealed coherent 401~Hz pulsations in the persistent X-ray flux, which 
directly arise from the neutron star rotation (Wijnands and van der Klis 1998).
SAX~J1808.4-3658 therefore represents the first accretion-powered 
millisecond X-ray pulsar. However kHz QPOs have not been observed yet 
from this source. 

In order to cover the range of frequencies observed,
we have chosen to study stars spinning with frequencies $290, 360$ and 
$580$~Hz. 
An uncertainty in the spin frequency of 30 Hz leads to an uncertainty
of $\sim 1 - 3$ Hz in the precession frequency (for two equal mass stars), 
the error increasing with the stiffness of the assumed EOS.  
We have also chosen stellar models spinning at a frequency of $720$ Hz,
although there exists no evidence that any of the low mass X-ray binaries
observed by RXTE are
spinning at this rate. (In fact this rate is faster than any of the 
known ms radio pulsars.) However, it is instructive to examine the effects of 
rapid rotation
on neutron stars and may be useful if faster neutron stars are observed in
the future. 

\subsection{Masses}

The optical lines of the mass donor stars in low-mass X-ray binaries 
are very difficult to observe, hence it is difficult to collect the 
radial velocities that are necessary to measure the neutron star masses. 
In only 
one Z source, Cyg X-2, it has been possible to measure the neutron star's
mass (Casares, Charles and Kulkeers 1998). A firm lower limit is
$1.27 M_\odot$. If it is assumed that the donor star has mass $> 0.75
M_\odot$, then the neutron star's mass is $> 1.88 M_\odot$. 
The mass of the neutron star in the transient low
mass X-ray binary Cen X-4 is still uncertain ($0.5-2.1 M_\odot$; 
Shahbaz et al. 1993).
The average mass of low mass X-ray binaries ({\it i.e.} 
neutron star plus companion star) inferred from their 
positions in globular clusters 
($\sim 1.5\pm^{0.4}_{0.6} M_\odot$; Grindlay et al. 1984) indicates 
a closer value to the canonical neutron star mass of $1.4 M_\odot$.

On the other hand 
the interpretation of the highest frequency of the 
kHz QPOs (especially in those sources in which this frequency 
is independent of the X-ray luminosity)
in terms of the orbital motion at the ISCO 
implies a neutron star mass in the $2.0-2.2\ M_{odot}$ 
range for several Atoll sources
(Kaaret, Ford and Chen 1997; Zhang, Strohamyer and Swank 1997; 
Zhang et al. 1998b). 

For each equation of state and for each angular frequency, at most
seven  different mass stars have been selected. Depending on the equation
of state and the spin rate, some of the stars described below will have 
the same mass, or may not be of interest.  
The lowest mass star  corresponds
to a star with a radius larger than the ISCO for both prograde
and retrograde orbits. The second and third lowest
mass stars have a radius slightly smaller (within $1\%$) than the ISCO
for retrograde and prograde orbits respectively. Other stellar models 
selected are as follows: a star with mass $1.4 M_\odot$; a star for which
the precession of prograde orbits at the ISCO is a maximum; a star with
a significant gap between the star's surface and the prograde ISCO.
The largest 
mass star in each case corresponds to the maximum mass normal sequence 
star allowed at that angular velocity. This is a star which, if spun down 
while keeping the number of baryons in the star constant, will have 
the same mass as the maximum mass non-rotating star. Higher mass stars,
called supramassive (Cook et. al. 1992) are possible, but at an angular
frequency of $360$ Hz, the fractional increase in mass is $<2\%$ for the 
stiffest equation of state (Cook et. al. 1994b).

\section{Numerical Results}
\label{s:general}

The results of the numerical computation of the precession frequencies for
corotating 
orbits inclined an infinitesimal angle to the star's equatorial plane
are now presented. For each stellar model selected by the criteria laid out
in the preceding section, the spacetime geometry and the worldlines of 
geodesics were computed using the formalism described in section 
\ref{s:models}. The equilibrium properties of these models are listed in
tables 1 - 4. For each model we have also listed the values of the
orbital frequency and the Lense-Thirring, and total relativistic precession
frequencies computed at the co-rotating ISCO for prograde orbits and at 
the counter-rotating ISCO for retrograd orbits. If the ISCO radius is 
smaller than the star's radius, its value is reported with the symbol 
``---'', and the frequencies reported are for orbits at the star's surface. 
 All values of radius reported in 
these tables refer to the Schwarzschild-like coordinate $r$ defined by 
Eq. (\ref{radius}).  
A discussion of the accuracy of these results will be 
postponed until the results have been presented. 

The key to tables 1 - 4 is as follows:

\begin{tabular}{ll}
$\nu_s$		& Star's spin frequency $\nu_s = \Omega/2 \pi$ [Hz]\\
$\bar{M}$	& Total mass, $\bar{M} = M/M_\odot$ \\
$\bar{M}_B$	& Baryon (or rest) mass, $\bar{M}_B = M_B/M_\odot$\\
$R$		& Radius of star, measured at the equator [km]\\
$I/\bar{M}$	
		& Ratio of moment of inertia to mass [$10^{45}$ g cm$^2$]\\
$\Phi_2$	& Quadrupole moment[$10^{43}$ g cm$^2$]\\
$j$		& Rotation parameter $cJ/GM^2$ \\
$\zeta$ 	& Centrifugal parameter (cf. Eq. (\ref{zeta})) \\
$r_{\pm}$	& Radius of innermost stable circular orbit [km] \\
$\nu_K$		& Kepler frequency at $r=r_{\pm}$ [kHz]\\
$\omega/2\pi$	
                & Lense-Thirring precession frequency at  $r=r_{\pm}$ [Hz]  \\
$\nu_{p}$	& Total precession frequency at $r=r_\pm$ [Hz]\\
\end{tabular}

%%%
These tables illustrate a number of features typical of the orbits of
particles at the minimum allowed orbital radius for a variety of masses 
and EOS. Two competing effects affect the precession frequency of a 
particle in a prograde orbit: the Lense-Thirring effect which scales as 
$j$ and is positive, and the precession due to oblateness, which scales
as $j^2$ and is negative. With this knowledge of the scaling, it can 
be understood how the the total prograde precession frequency varies 
with $\bar{M}$ and $\nu_s$. If $\nu_s$ is held fixed and $\bar{M}$ allowed to
vary, $j=J/M^2$ will increase quadratically, causing the precession due to 
oblateness to grow rapidly. The overall effect will be to reduce the total
precession. If $\bar{M}$ is held fixed while $\nu_s$ increases, $j$ increases
linearly. As long as $j$ is small, ($j\leq 0.2$) a small linear increase
in $j$ will tend to increase the Lense-Thirring precession while the 
oblateness precession will not change very much. Hence, we expect that the 
total precession will increase. Once $j$ is larger than $\sim 0.2$, the
quadratic dependence of the oblateness precession will cause $\nu_p$ to 
decrease with increasing $j$. This general behaviour is observed in the 
models presented in the tables which have radius larger than the ISCO 
radius ($R> r_+$).
%%%

The quadrupole moments listed in these tables are computed using the 
method described by Laarakkers and Poisson (1997) and agree with their
results when models overlap.
The quadrupole moments scale roughly as the 
square of $\nu_s$ for stars of constant mass for slow rotation. Quadrupole
moments for stars spinning at a rate different from those listed in the 
tables can be found by finding the quadrupole moment for a star with the 
same mass and scaling appropriately. The quadrupole moment for a star with
mass different from the masses listed in the tables in the present paper
or that of Laarakkers and Poisson can be found by interpolation.

\subsection{Dependence on orbital frequency}

%LS
Within beat frequency models, the observed variations in the  
centroid frequency of the kHz QPOs in a given source, correspond
to changes in radial position
of the inner edge of the accretion disk around the neutron star. 
As the radius changes, the 
orbital and precession frequencies will change in a correlated fashion. 
If the precession is dominated by the L-T effect, then 
the precession should
vary as the square of orbital frequency (Stella and Vietri 1998a). Close 
to the star relativistic effects may be important and large deviations from 
the quadratic behaviour may occur. 

\begin{figure}
\plotone{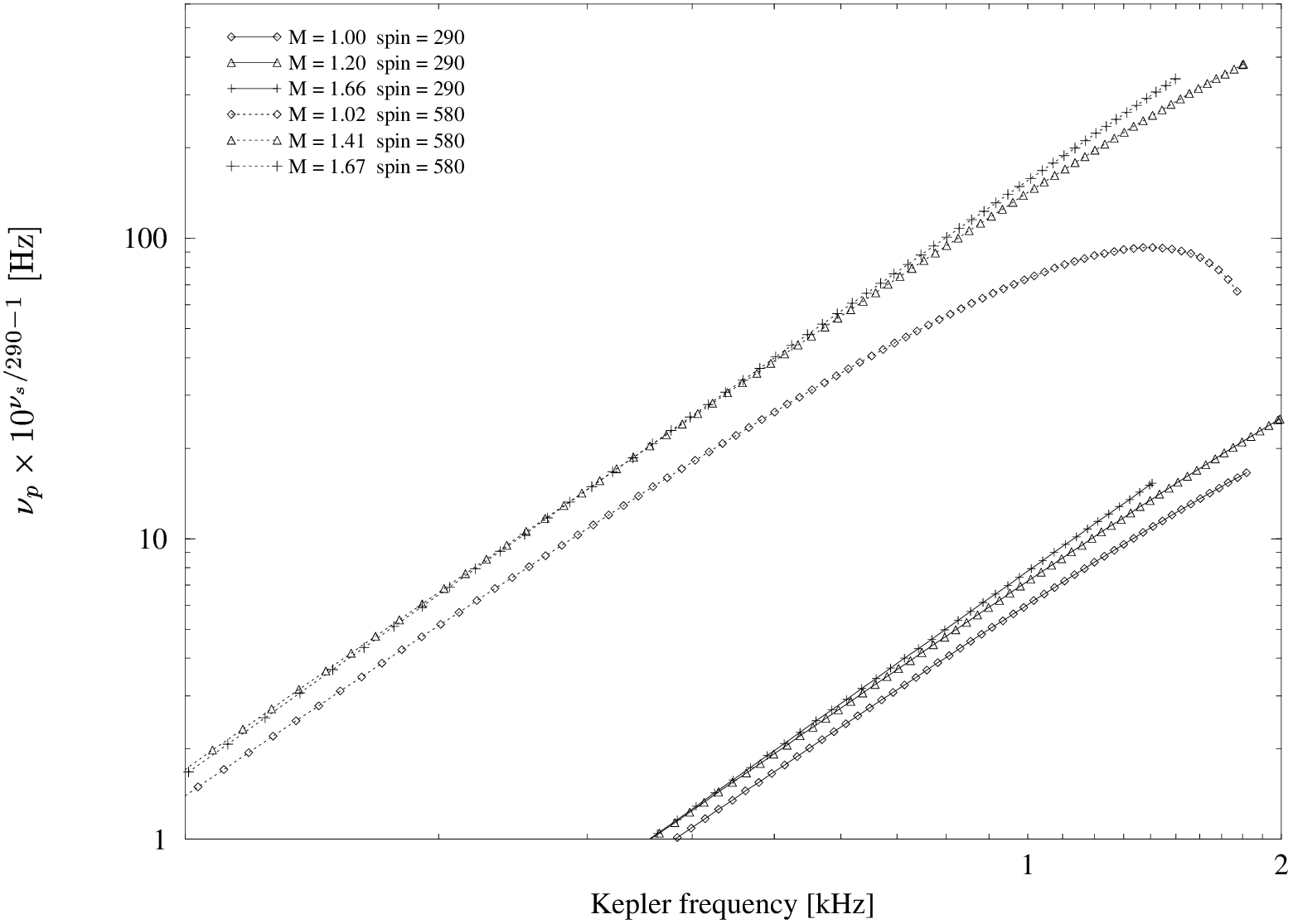}
\caption{EOS A: Precession frequency versus orbital (Kepler) 
frequency for stars
with spin frequency $\nu_s = 290, 580$ Hz. 
Each curve corresponds to a star with 
different mass. For clarity, the precession frequencies for stars 
rotating at a 
frequency of 580 Hz have been multiplied by a factor of 10.  \label{graphA1}}
\end{figure}

\begin{figure}
\plotone{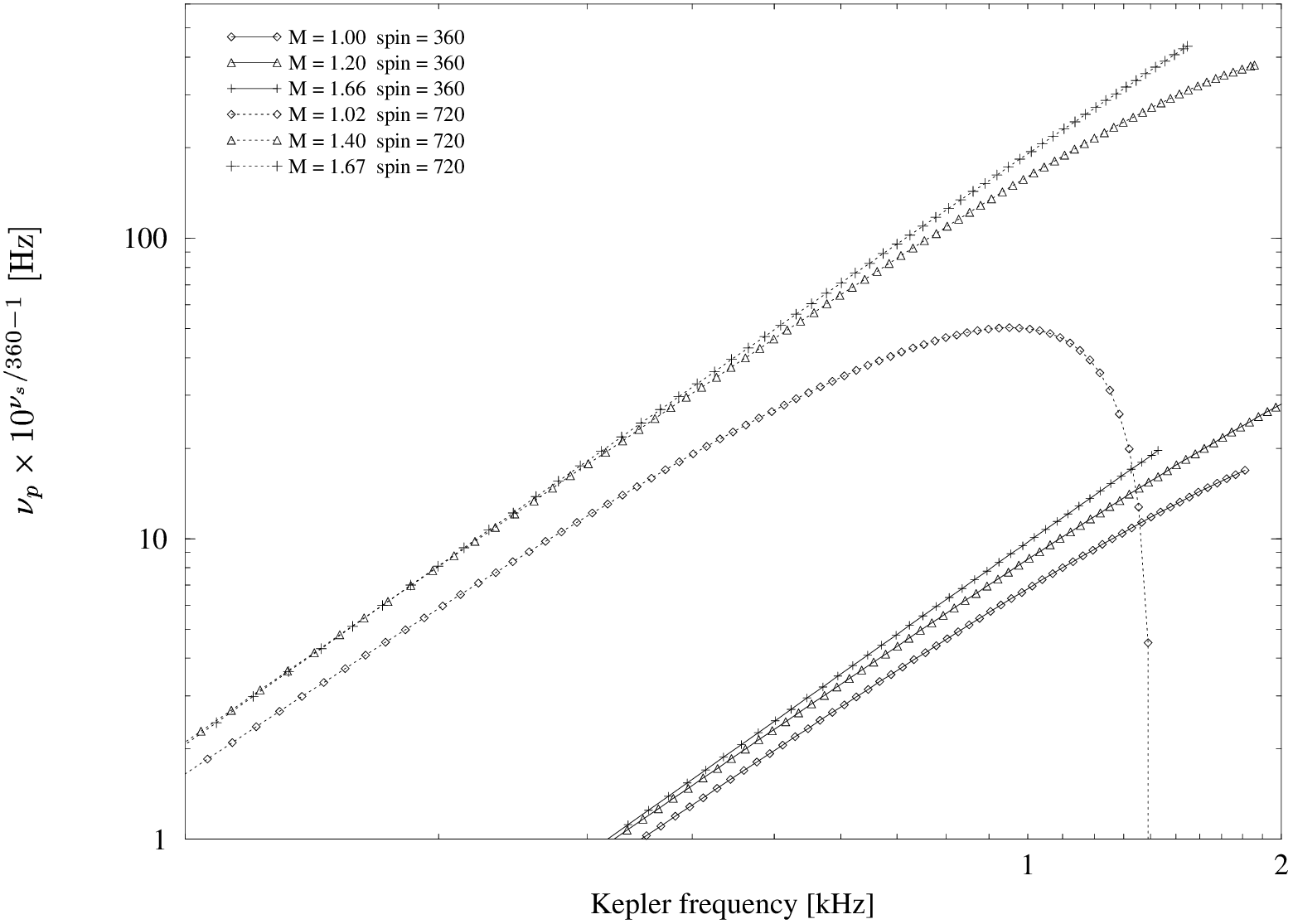}
\caption{EOS A: Precession frequency versus orbital (Kepler) 
frequency for stars
with spin frequency $\nu_s = 360, 720$ Hz. 
Each curve corresponds to a star with 
different mass. For clarity, the precession frequencies for stars 
rotating at a 
frequency of 720 Hz have been multiplied by a factor of 10.  \label{graphA2}}
\end{figure}

\begin{figure}
\plotone{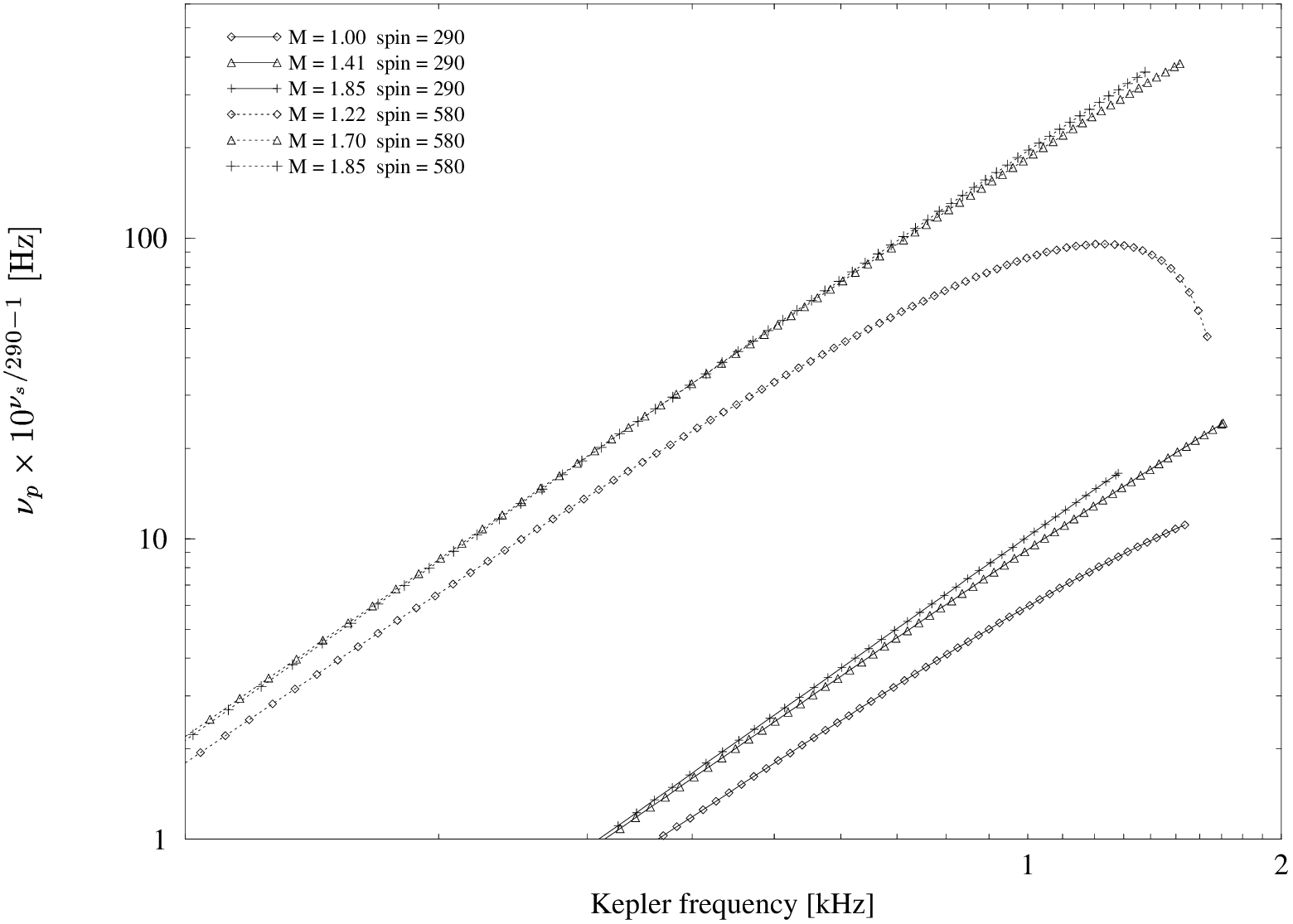}
\caption{EOS WFF3: Precession frequency versus orbital (Kepler) 
frequency for stars
with spin frequency $\nu_s = 290, 580$ Hz. 
Each curve corresponds to a star with 
different mass. For clarity, the precession frequencies for stars 
rotating at a 
frequency of 580 Hz have been multiplied by a factor of 10.  \label{graphW1}}
\end{figure}

\begin{figure}
\plotone{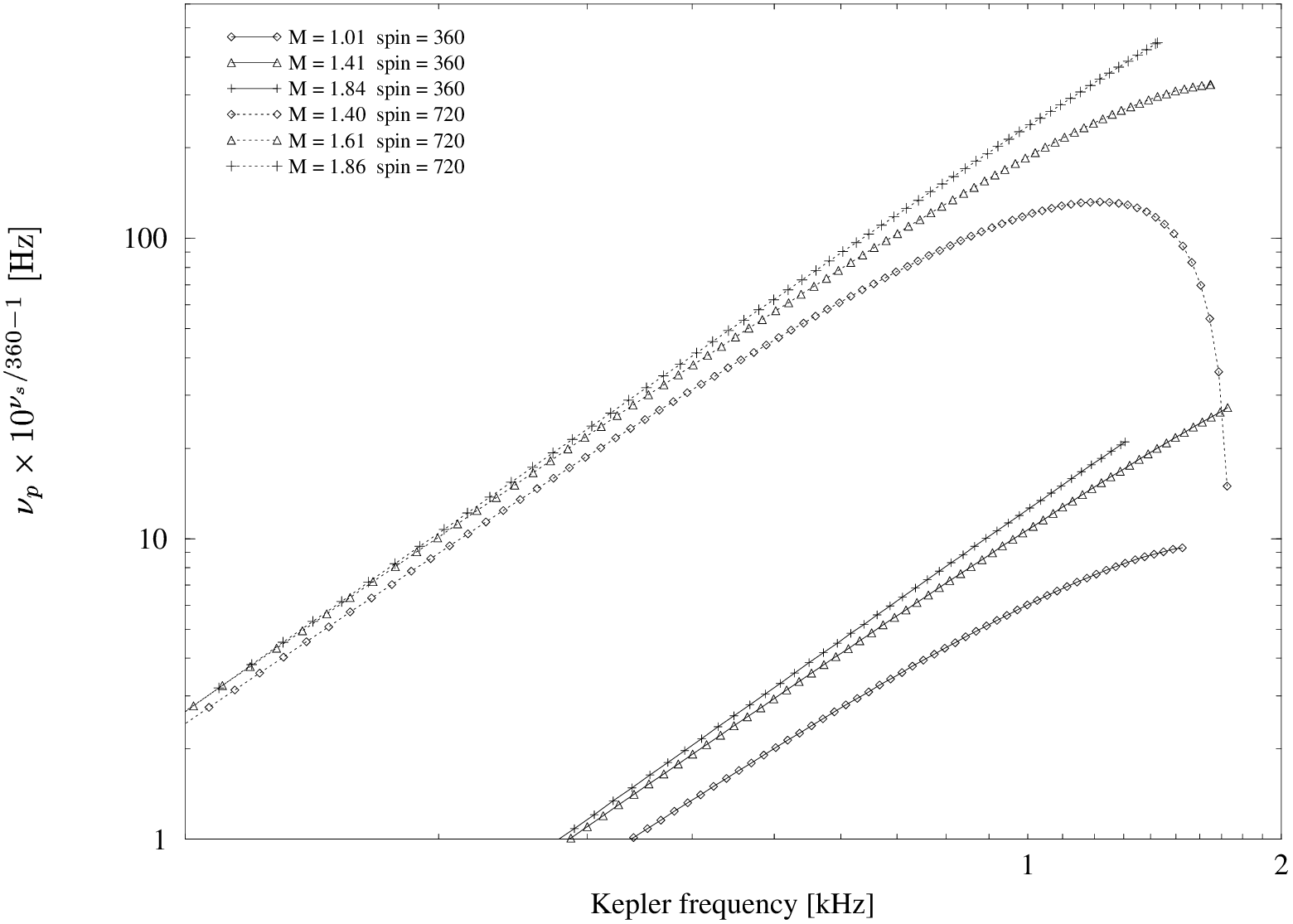}
\caption{EOS WFF3: Precession frequency versus orbital (Kepler) 
frequency for stars
with spin frequency $\nu_s = 360, 720$ Hz. 
Each curve corresponds to a star with 
different mass. For clarity, the precession frequencies for stars 
rotating at a 
frequency of 720 Hz have been multiplied by a factor of 10.  \label{graphW2}}
\end{figure}

\begin{figure}
\plotone{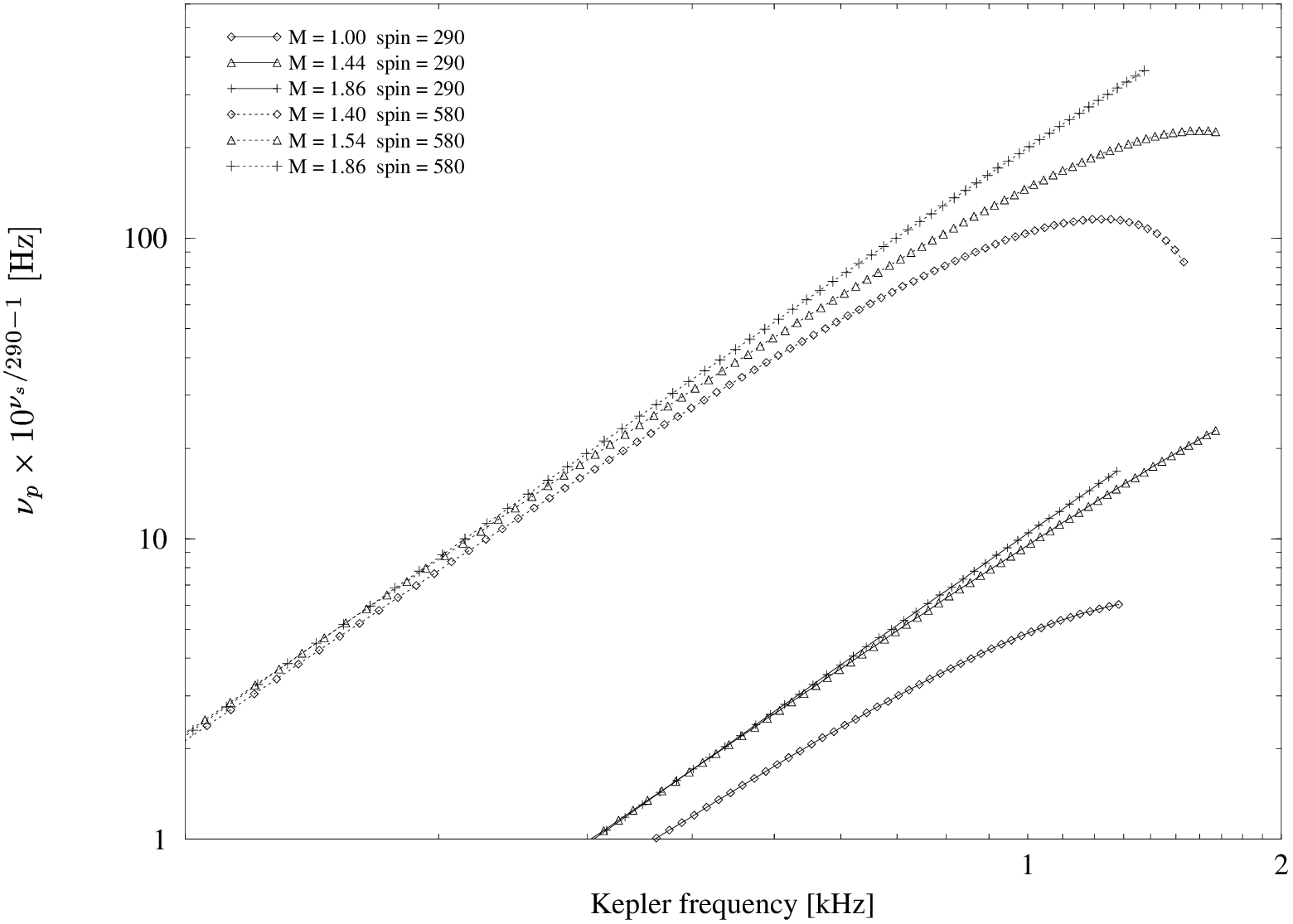}
 
\caption{EOS C: Precession frequency versus orbital (Kepler) 
frequency for stars
with spin frequency $\nu_s = 290, 580$ Hz. 
Each curve corresponds to a star with 
different mass. For clarity, the precession frequencies for stars 
rotating at a 
frequency of 580 Hz have been multiplied by a factor of 10.  \label{graphC1}}
\end{figure}

\begin{figure}
\plotone{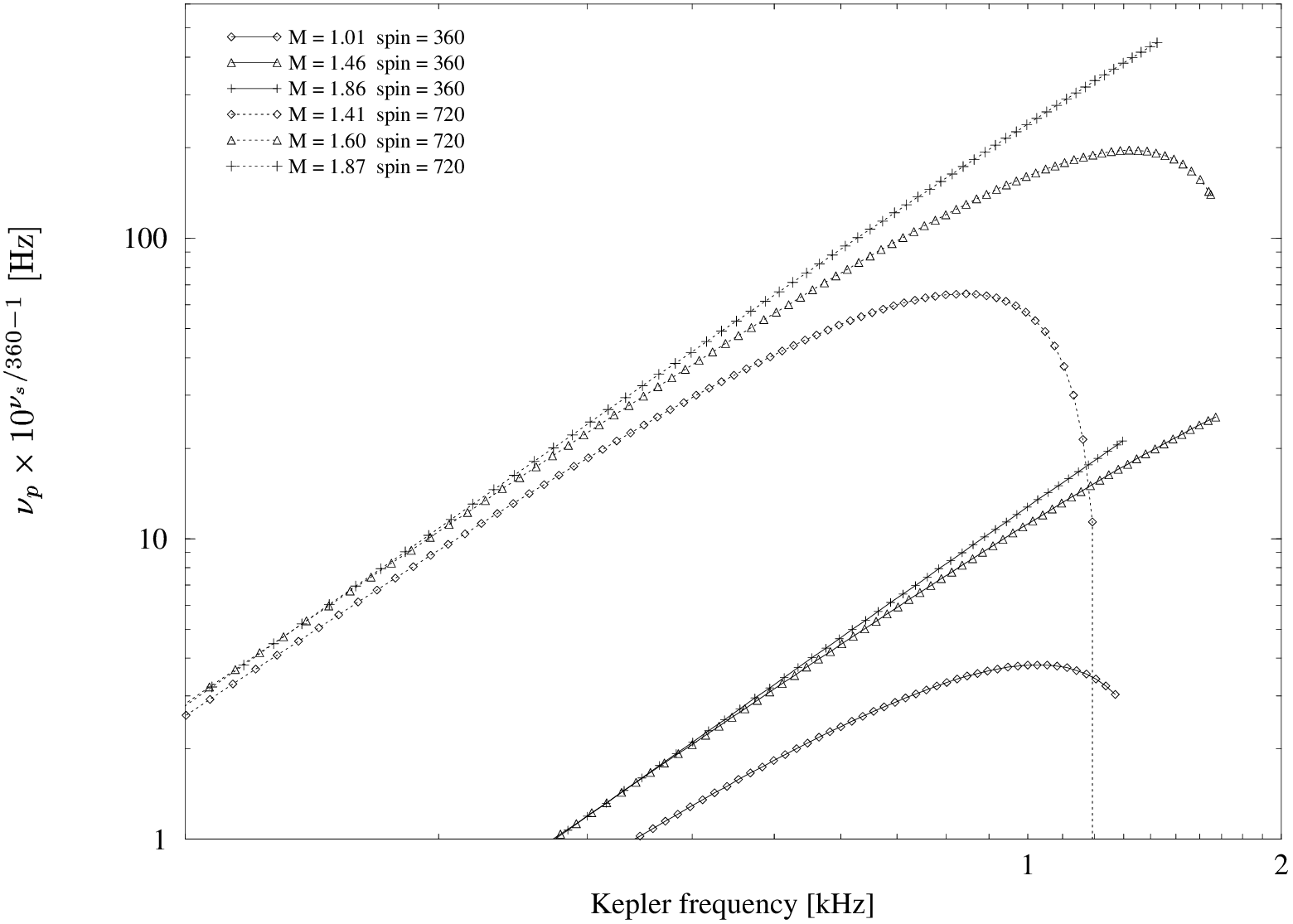}
 
\caption{EOS C: Precession frequency versus orbital (Kepler) 
frequency for stars
with spin frequency $\nu_s = 360, 720$ Hz. 
Each curve corresponds to a star with 
different mass. For clarity, the precession frequencies for stars 
rotating at a 
frequency of 720 Hz have been multiplied by a factor of 10.  \label{graphC2}}
\end{figure}

\begin{figure}
\plotone{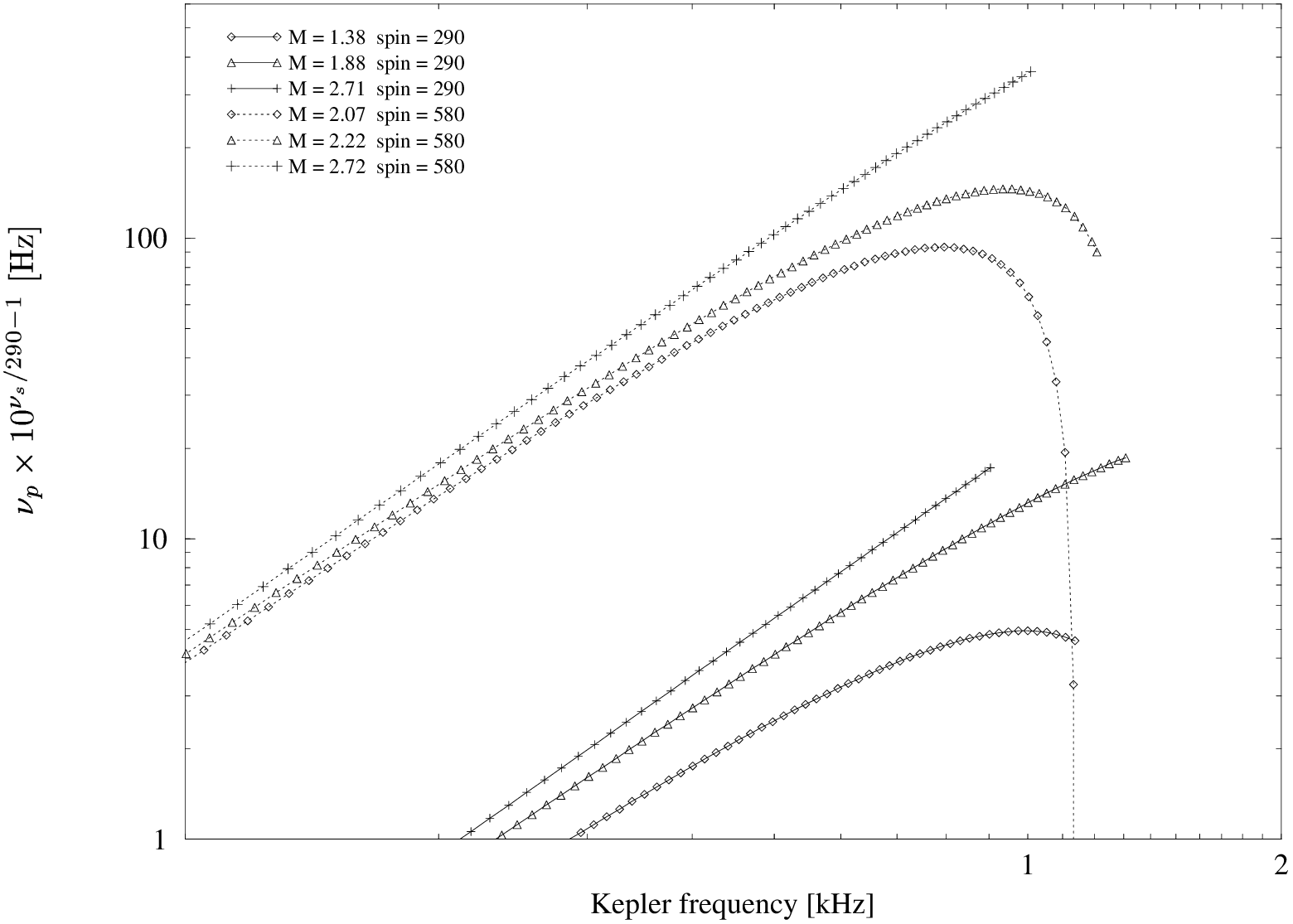}
 
\caption{EOS L: Precession frequency versus orbital (Kepler) 
frequency for stars
with spin frequency $\nu_s = 290, 580$ Hz. 
Each curve corresponds to a star with 
different mass. For clarity, the precession frequencies for stars 
rotating at a 
frequency of 580 Hz have been multiplied by a factor of 10.  \label{graphL1}}
\end{figure}

\begin{figure}
\plotone{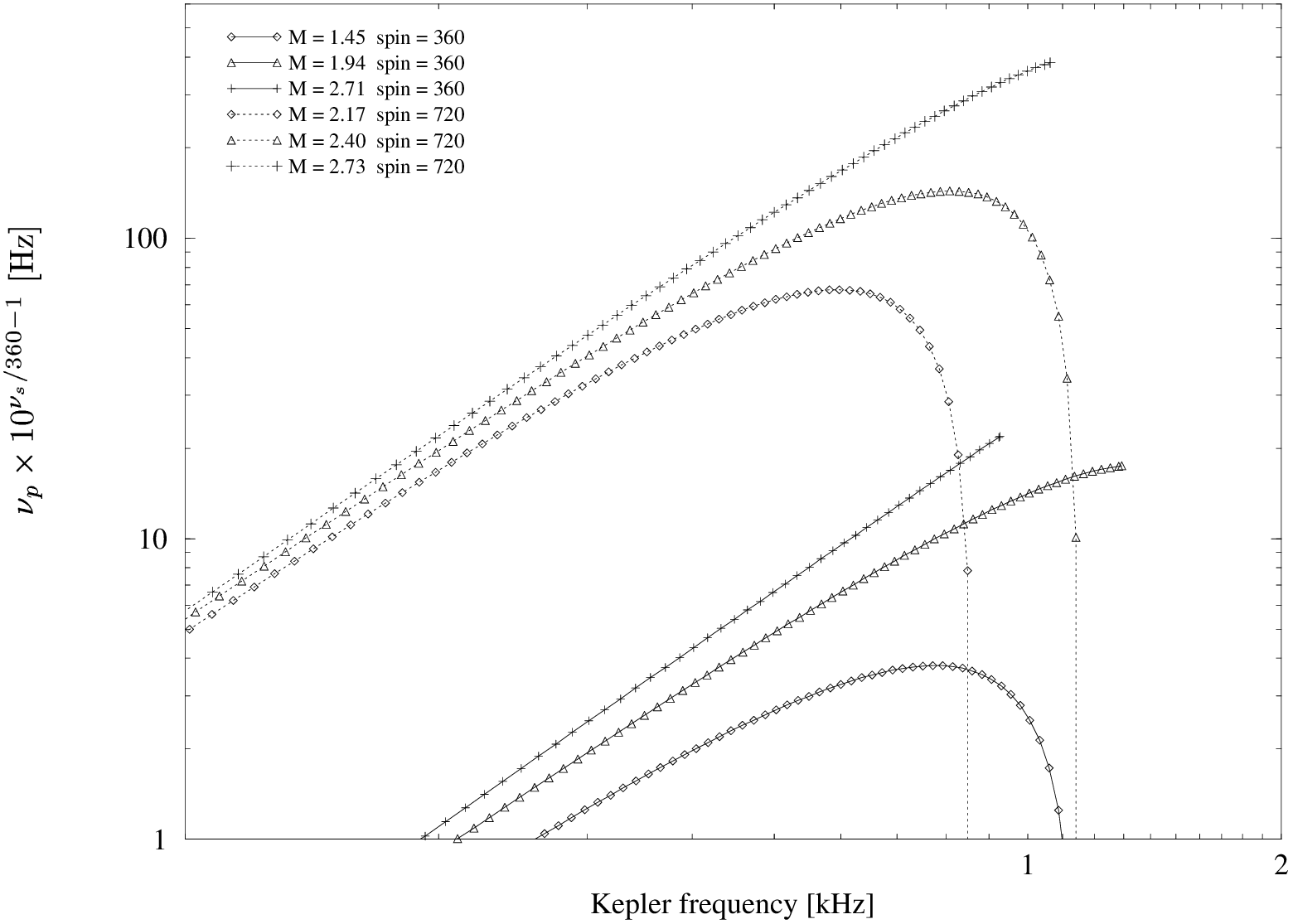}
\caption{EOS L: Precession frequency versus orbital (Kepler) 
frequency for stars
with spin frequency $\nu_s = 360, 720$ Hz. 
Each curve corresponds to a star with 
different mass. For clarity, the precession frequencies for stars 
rotating at a 
frequency of 720 Hz have been multiplied by a factor of 10.  \label{graphL2}}
\end{figure}

In Figures \ref{graphA1} - \ref{graphL2} graphs of precession frequency 
versus Kepler (or orbital) frequency are plotted for a subset of the 
stars featured in tables 1 - 4. Each graph features a number of 
stars with two different angular frequencies, the more rapidly 
rotating  stars (dotted
lines) shifted upwards by a factor of 10 for clarity. The frequencies 
are plotted log-log to make it easy to discern 
%LS
departures from a power-law behaviour. 
Each curve corresponds to a different stellar model. Each point on the
curve corresponds to a different radius at which the precession 
and 
orbital frequencies are evaluated. 

The termination of each curve corresponds to the smallest radius (or
highest orbital frequency) at which a particle can orbit the star. This 
limiting radius is either the surface of the star or the innermost stable
circular orbit. 

As an example of the use of these graphs, 
consider the power spectra displayed
in Figure 3 of Strohmayer et. al. (1996) for the Atoll source 4U 1728-34, 
the neutron star of which rotates at $\nu_s \sim 360$~Hz.
Broad peaks with centroids near 20, 26 and 35 Hz are clearly seen, which
Stella and Vietri (1998a) propose to interpret as due to L-T precession. 
At the 
same time, the higher kHz QPOs,  
interpreted as the orbital frequency, are observed at $\sim 900$, 980 and 
1100 Hz respectively.  
These data points do not lie along any of the curves on 
Figures \ref{graphA2}, \ref{graphW2}, \ref{graphC2} or \ref{graphL2}. 
 
%%%
Although the observational data for 4U1728-34 do not coincide with any of 
the theoretical curves, it can be seen that stellar models exist for which
the observed QPOs are close to twice the theoretical precession frequencies.
In table 5 we summarize the masses for each EOS which come closest to having
precession frequencies  which are one half the observed frequencies. 
We note that for EOS WFF3 and C, masses in the range $1.6M_\odot - 1.86
M_\odot$ provide the closest fits. For EOS L, a mass near $2.0 M_\odot$ comes
closest to producing precession frequencies near one half the observed values.
It should be noted that for EOS L, the general relativistic effects produce
a considerably smaller precession frequency than was predicted from 
the semi-Newtonian analysis done by Stella and Vietri (1998a).
%%%

Data from other atoll sources show a similar trend. For example, the 
source KS 1731-260   (Wijnands and van der Klis 1997) spinning at either
$260$ or $520$ Hz, shows a broad feature centred about 27 Hz
when the higher frequency kHz QPO is 
at 1200 Hz. 
Again, the predicted 
precession frequencies  are close to half of what is observed,
if the star is spinning with a frequency of 260 Hz. However, it the star
is spinning at 520 Hz, then the theoretical precession frequencies 
are close to the observed QPO. In table 6 we summarize the masses
which provide the closest fits to the data for KS 1731-260, assuming
the neutron star is spinning with a frequency of 260 or 520 Hz.

The situation for the Z sources is a little different. These sources exhibit
a somewhat more coherent peak in the $10 - 60$ Hz range, which is called the 
horizontal branch oscillation (HBO). For example, GX 17+2 (Wijnands et. al.
1997b) has a HBO peak at 60 Hz while the highest QPO is at 1000 Hz. This 
star is spinning at 290 Hz. The precession 
frequencies predicted for the softer EOS are of order 10 Hz for an orbital
frequency of 1000 Hz, which a factor of $\sim 5-6$ too small to explain the 
observations. EOS L allows a precession of 20 Hz if the star is very heavy,
$\sim 2.4 M_\odot$, but this is still a factor of 3 too small. 
If the star is rotating at $580$ Hz,
the predicted precession rate is still too low: EOS C predicts a
precession rate of only 20 Hz for the  $1.8 M_\odot$ star.
%LS
It seems unlikely that simple precession models explain the 
HBO peaks seen in the Z sources.

\subsection{Accuracy of the post-Newtonian formula}

We will now test the validity of a simplified post-Newtonian 
precession formula in the slow rotation limit. For simplicity we will 
consider a formula which ignores the centrifugal precession and the $O(j^2)$
post-Newtonian correction to the quadrupole moment. 
The approximate formula
is 
\be
\nu_p \simeq \nu_{LT} - \nu_{quad}(1 + 3\frac{GM}{rc^2}) , 
	\label{approx_prec}
\ee
where the Lense-Thirring precession is
\be
\nu_{LT} \simeq  13.2 \;\frac{I}{\bar{M}}\;
	\nu_{s,2.5}  
	\left( \nu_{K,3} \right)^2 \ Hz
	\label{approx_lt}
\ee
and the Newtonian quadrupole precession is 
\be 
\nu_{quad} \simeq 3.5 \;\Phi_{2,43}\;
	\left( \bar{M} \right)^{-5/3}
	\left( \nu_{K,3} \right)^{7/3}\ Hz .
	\label{approx_quad1} 
\ee
In Eqs. (\ref{approx_lt}) and (\ref{approx_quad1}) we have defined a number
of scaled variables defined by: $\bar{M} = M/M_\odot$, $I_{45} = I/10^{45}$,
$\Phi_{2,43} = \Phi_2/10^{43}$, $\nu_{s,2.5} = \nu_s/300$ and
$\nu_{K,3} = \nu_K/10^3$.

In terms of the form factors $a(M,\hbox{EOS})$ given by 
Laarakkers and Poisson (1997),
the Newtonian quadrupole precession is
\be
\nu_{quad} \simeq 6.5 \frac{a(M,\hbox{EOS})}{10}
	\left(\frac{I}{\bar{M}} \;\nu_{s,2.5}\right)^2
	\left(  \bar{M} \right)^{-2/3}
	\left( \nu_{K,3} \right)^{7/3}\ Hz.
	\label{approx_quad2}
\ee
The post-Newtonian correction term is 
\be
3\frac{GM}{rc^2} \simeq  0.3 \left( \bar{M} \right)^{2/3}
	\left( \nu_{K,3} \right)^{2/3}.
	\label{approx_pn}
\ee

Consider, as an example, a moderately slowly rotating star such as the
$M=1.0 M_\odot$, $\nu_s = 290$ Hz star described by EOS C. 
 The surface of this star is the minimal radius
allowed for both prograde and retrograde orbits. The formula 
(\ref{approx_lt}) predicts a precession due to the Lense-Thirring effect
of 20 Hz, given data from the first row of Table \ref{t:tableC}.
From the definition given by Eq. (\ref{lt_freq}), we find that 
$\nu_{K+} + \nu_{K-} = \omega/2 \pi = 20 Hz$, as given by the slow
rotation formula. The average of the prograde and retrograde precession 
frequencies, $\frac{1}{2} (\nu_{p+} + \nu_{p-}) = 19 Hz$ also agrees 
with $\omega/2\pi$ within $0.5\%$. The precession due to oblateness 
Eq. (\ref{oblate_freq})  is 
$\nu_{oblate} = \frac{1}{2}(\nu_{p-} - \nu_{p+}) = 13 Hz$. But the 
Newtonian quadrupole precession given by Eq. (\ref{approx_quad1}) is only 
10 Hz. Inclusion of the post-Newtonian correction Eq. (\ref{approx_pn}) 
increases this to 13.5 Hz
corresponding to $4\%$ error. Clearly the post-Newtonian 
correction must be included close to the star. Finally we note that the
centrifugal precession frequency is only $1\%$ of the quadrupole frequency
(as given by the parameter $\zeta$ in the tables). Hence this frequency
is small enough that it can be neglected. Since the parameter $\zeta$ 
reaches large values $\sim 10 \%$ only when the quadrupole moment is small
it seems likely that the centrifugal precession can always be ignored at 
this level of approximation.  

\begin{figure}
\plotone{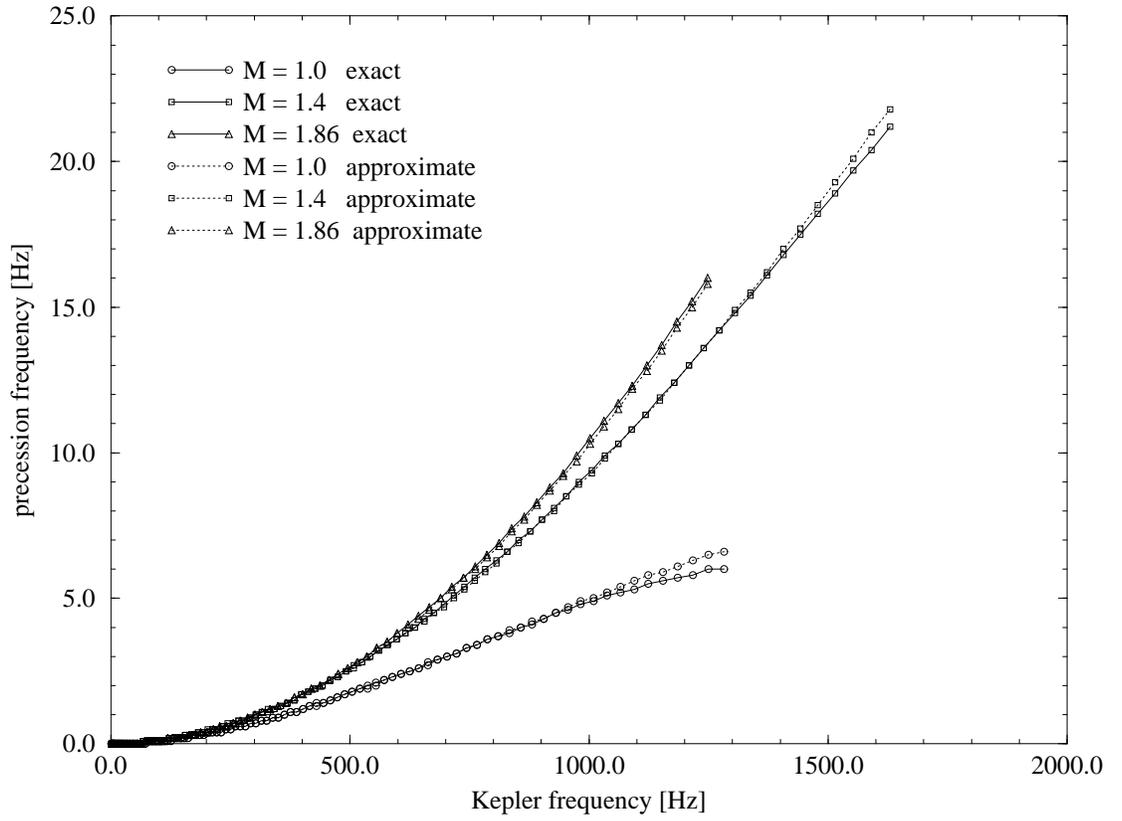}
\caption{Comparison of post-Newtonian and exact precession frequencies for
EOS~C and $\nu_s = 290$~Hz \label{approx}}
\end{figure}

In Figure \ref{approx} a comparison of the approximate 
post-Newtonian formula for
the precession of prograde orbits with the exact results is made. Prograde
precession frequencies for three 
EOS C stars with masses $1.0, 1.4$ and $1.86 M_\odot$, 
rotating with $\nu_s = 290$ Hz are plotted versus orbital frequency. 
In each case the
simplified post-Newtonian formula (\ref{approx_prec})
is plotted as a dotted line, while the exact precession is plotted as 
a solid line. At orbital frequencies $\nu_K < 1000$ Hz, 
the exact and approximate curves
are indistinguishable. All of these models have $j<0.2$, and in this 
sense are slowly rotating. The precession frequency for stars with 
$j>0.2$ won't be approximated very 
well by Eq. (\ref{approx_prec}) and must be computed numerically.

\subsection{Numerical accuracy}

The numerical solutions which we find use a code (Stergioulas 1995) 
based on that of Cook et. al. (1994). The numerical grid is composed of
201 points in the radial direction and 101 points in the angular direction 
which is a finer grid than used by Cook et. al. This finer grid size was 
required to keep the errors in the numerically computed precession frequencies
down to $1\%$. 
In the series expansions of the metric functions given by Eqs. (\ref{gamma}) - 
(\ref{omega}), all terms up to and including $n=10$ were used. We checked 
that our numerical values of masses and radii agreed within $1\%$ of those
found by Cook et. al. for EOS A and L.

\begin{figure}
\plotone{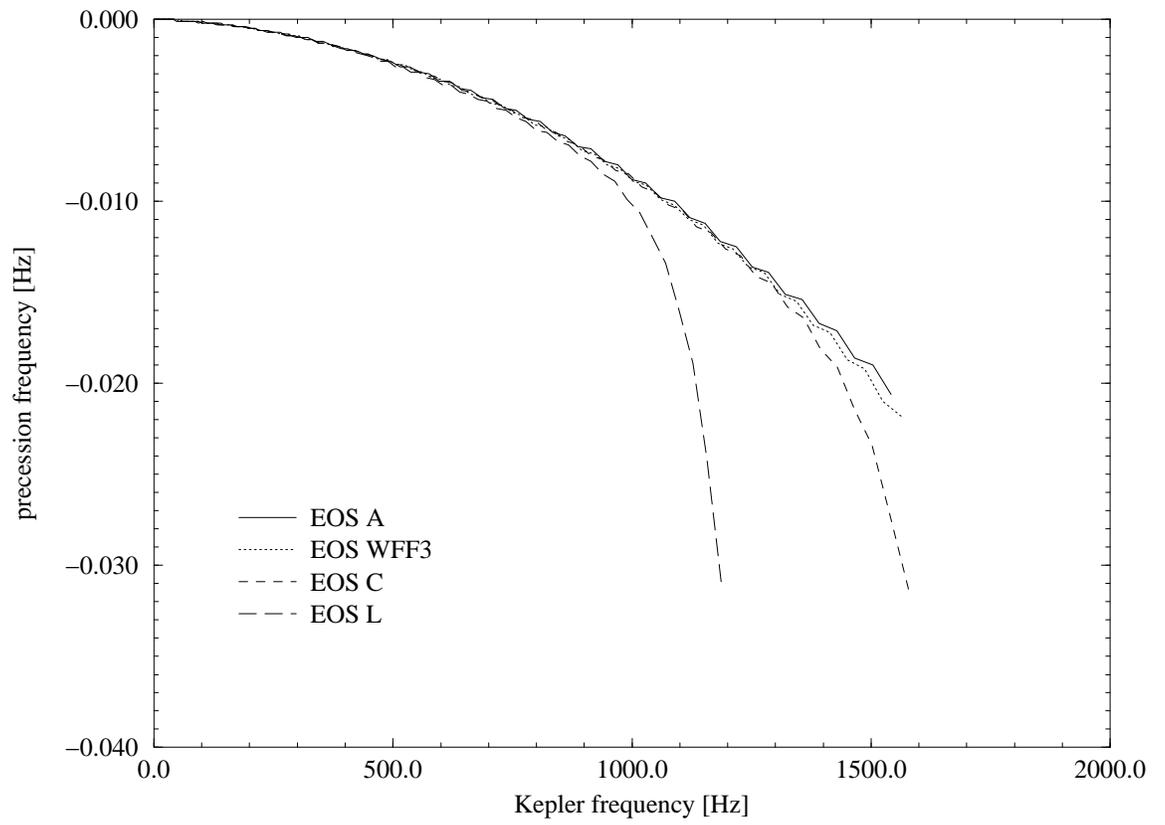}
\caption{Precession frequencies for non-rotating stars with $M= 1.4 M_\odot$
 \label{errors}}
\end{figure}

To find the size of the absolute errors, we computed the precession for
non-rotating stars with $M=1.4 M_\odot$, 
the results of which are displayed in Figure 
\ref{errors}.  Theoretically, the precession of orbits is zero if the 
star is static. The numerically computed 
precession was less than $0.03$ Hz. The precession fell to less than $0.01$ at
an orbital frequency of $1000$ Hz. These small but non-zero frequencies give
a measure of the absolute numerical error inherent to our method. 

%%%
A number of tests were done to check the validity of the frequencies found.
Far from the star ($r/M > 10$) one half the sum of the prograde and
retrograde precession frequencies at constant radius must reduce to the
standard expression for the Lense-Thirring precession,
\be
\frac12 \left( \nu_{p+} + \nu_{p-} \right) = \frac{2J}{r^3}
	\left( 1 + O(j^2) \right).
\label{sum}
\ee
Similarly, the difference must reduce to the classical quadrupole precession,
\be
\frac12 \left( \nu_{p+} - \nu_{p-} \right) = - \frac32 \nu_K
	\frac{\Phi_2}{Mr^3} \left( 1 + O(j^2) \right).
\label{difference}
\ee
The quantities $J$ and $\Phi_2$ can be computed independently of $\nu_{p\pm}$.
Expressions (\ref{sum}) and (\ref{difference}) were satisfied to better than
$1\%$ for large radii. The values of the precession frequencies in the
strong field region near the ISCO were validated by noting that the expression
on the left-hand-side of equation (\ref{sum}) should differ from the 
metric function $\omega/2\pi$ only at order $j^3$. (The computation of 
the precession frequencies depends on $\omega$ in a non-linear fashion
through Eqs. (\ref{precession}) and (\ref{x}), 
so this is a non-trivial comparison.) Hence the inequality
\be
\frac{\pi \left( \nu_{p+} + \nu_{p-} \right) - \omega}{\omega}
	\leq O(j^2)
\label{inequality}
\ee
must be satisfied. We found that the inequality (\ref{inequality})
was satisfied for the models presented in this paper. When the value 
of the roation parameter took values $j\leq 0.2$, the left-hand-side
of the inequality (\ref{inequality}) was smaller than $j^2$, while it was 
larger than $j^2$  but of the same order of magnitude as $j^2$ when 
$j>0.2$. We conclude then that the precession frequencies computed are correct
to within $1\%$.

\section{Conclusion}
\label{s:conclusion}

We have computed the dependence of precession frequency on orbital frequency
for a wide variety of neutron star masses, spins and equations of state. 
In doing so we have made a number of simplifying assumptions about the 
physics of the region close to the neutron star. The main assumptions
used in this paper are:
\begin{enumerate}
\item Some physical mechanism exists to tilt the inner orbits of the 
accretion disk out of the star's equatorial plane. 
\item All forces on particles in the disk's inner region besides gravity
are negligible.
\item The perturbations causing the tilt create only an infinitesimal 
tilt angle. 
\end{enumerate}
We now consider the validity and ramifications of these assumptions. 

It has long been thought that the combination of viscosity and 
gravito-magnetism always act to keep the inner ($r < 100 M$) region of
an accretion disk co-planar with the star (Bardeen and Petterson 1975). 
However, recent calculations have shown that it is possible for warped 
precessing disk modes to exist in the inner disk region. Ipser (1996) 
applied perfect fluid perturbation theory to show that such modes are 
possible. The modes which he finds for the Kerr black hole metric are
fairly low, $\leq 10Hz$. Adopting the formalism 
of Papaloizou and Pringle (1983), Markovi\'{c} and Lamb (1998) have included 
viscosity in the computation and have shown that a family of highly 
localised and weakly damped  tilt modes exist close to the inner 
disk boundary which precess at 
a frequency very close to the local precession frequency. 
Alternatively it has been demonstrated that if the accretion disk matter is 
inhomogeneous, diamagnetic blobs can be lifted above the equatorial plane 
(where they will start precessing) 
through the resonant interaction with the star's magnetic field  
(Vietri and Stella 1998). 

Our assumption that gravity is the only force acting on particles in the 
inner region will not always be valid. Miller (1998) has shown that the
interaction of radiation emitted close to the neutron star surface 
with the disk particles can have an important 
effect on the precession frequency, increasing it by a large factor. 
The effect is most pronounced for high luminosity sources (approaching the 
Eddington limit, such as Z-sources). 
However, it is unclear that the approach discussed by Miller (1998) 
is applicable 
to warped disks of high Thomson depth. In the framework of inhomogeneous 
accretion disks, the effects of radiation drag on individual diamagnetic blobs 
are probably negligible (Vietri and Stella 1998). 
In any case, the frequencies calculated in this paper should be treated 
as baseline frequencies for neutron stars which might be modified by 
the inclusion of other physics. 

It seems reasonable to expect that the physical mechanisms explored so far
will only produce a small tilt out of the equatorial plane. If,
however, the tilt angle is large, the present results can nevertheless 
be used to 
estimate the precession frequency. To do so, we recall that the 
Lense-Thirring frequency is unaffected by the tilt angle, while the
Newtonian quadrupole precession is multiplied by a factor of $\cos \beta$,
where $\beta$ is the angle between the disk and the star's equatorial plane. 
Inclusion of this factor in the approximate post-Newtonian formula 
(\ref{approx_prec}) will provide a good estimate of the precession of 
orbits tilted by a large angle.

Finally, we note that the precession frequencies which 
we compute are a factor of $\sim 2$ smaller than the low-frequency
($\sim 20-35$~Hz)
QPOs observed in the Atoll sources, that Stella and Vietri (1998a) 
interpreted in terms of precession frequency. 
In order for precession to explain
these results, a physical or geometric mechanism to produce QPOs at twice the 
theoretical frequency must be developed. A possibility is that a modulation at 
twice the precession frequency is generated at the two points where the 
inclined orbit of diamagnetic blobs intersect the disk 
(Vietri and Stella 1998). 
   
\acknowledgments
We would like to thank John Friedman, Eric Poisson and Mario Vietri 
for many useful 
discussions. We would especially like to thank 
Nikolaos Stergioulas for making his computer code publicly
available. This work was supported in part by NSERC of Canada, NSF grant
PHY 95-07740
%LS
and ASI grants.

%%%%%%%%%%%%%%%%%%%%%%%%%%%%%%%%%%%%%%%%%%%%%%%%%%%%%%%%%%%%%%%%%%%%%%%%%%%%%

\clearpage

\clearpage

\begin{table}
\label{t:tableA}
\caption{Precession frequencies, EOS A}

\begin{tabular}{|rrrrrrr|rrrr|rrrr|}\tableline
\multicolumn{7}{|c|}{ } &
\multicolumn{4}{c|}{prograde} &
\multicolumn{4}{c|}{retrograde}\\
\tableline

$\bar{M}$	&$\bar{M}_B$	&$R$	&$I/{\bar{M}}$	&$\Phi_2$	&$j$	&$\zeta$ 
		&$r_{+}$	&$\nu_K$	&$\omega/2\pi$	&$\nu_p$ 
		&$r_{-}$	&$\nu_K$	&$\omega/2\pi$	&$\nu_p$ \\

\tableline
\multicolumn{15}{|c|}{ $\nu_s = 290$ Hz } \\
\tableline
	1.00 &	1.08 &	10.0 &	 0.7 &	 0.6 &	0.14 &	   2 &	 --- &	1.82 &	28.3 &	16.6 &	 --- &	-1.85 &	28.3 &	38.1 \\
	1.05 &	1.15 &	10.0 &	 0.7 &	 0.6 &	0.13 &	   2 &	 --- &	1.88 &	30.9 &	19.2 &	10.1 &	-1.86 &	29.3 &	38.4 \\
	1.20 &	1.32 &	 9.9 &	 0.7 &	 0.5 &	0.12 &	   3 &	10.0 &	1.99 &	36.0 &	25.0 &	11.4 &	-1.67 &	24.4 &	29.9 \\
	1.41 &	1.59 &	 9.6 &	 0.7 &	 0.4 &	0.11 &	   5 &	11.8 &	1.69 &	26.3 &	21.8 &	13.2 &	-1.44 &	18.7 &	21.3 \\
	1.61 &	1.87 &	 9.1 &	 0.7 &	 0.3 &	0.09 &	  10 &	13.5 &	1.46 &	19.4 &	17.6 &	14.9 &	-1.28 &	14.5 &	15.5 \\
	1.66 &	1.95 &	 8.4 &	 0.6 &	 0.2 &	0.08 &	  15 &	14.1 &	1.41 &	16.3 &	15.3 &	15.4 &	-1.25 &	12.6 &	13.2 \\
\tableline
\multicolumn{15}{|c|}{ $\nu_s = 360$ Hz } \\
\tableline
	1.00 &	1.08 &	10.1 &	 0.7 &	 0.9 &	0.17 &	   2 &	 --- &	1.81 &	35.2 &	16.9 &	 --- &	-1.85 &	35.2 &	49.9 \\
	1.03 &	1.12 &	10.0 &	 0.7 &	 0.9 &	0.17 &	   2 &	 --- &	1.85 &	37.0 &	18.7 &	10.2 &	-1.83 &	35.2 &	48.8 \\
	1.20 &	1.33 &	 9.9 &	 0.7 &	 0.8 &	0.15 &	   3 &	 9.9 &	2.01 &	46.2 &	28.4 &	11.6 &	-1.62 &	28.7 &	36.3 \\
	1.41 &	1.60 &	 9.6 &	 0.7 &	 0.7 &	0.13 &	   5 &	11.7 &	1.71 &	34.1 &	26.7 &	13.4 &	-1.41 &	22.3 &	25.9 \\
	1.61 &	1.87 &	 9.1 &	 0.7 &	 0.5 &	0.11 &	  10 &	13.4 &	1.49 &	24.9 &	22.1 &	15.1 &	-1.26 &	17.3 &	18.8 \\
	1.66 &	1.95 &	 8.5 &	 0.6 &	 0.3 &	0.10 &	  15 &	13.9 &	1.43 &	21.3 &	19.6 &	15.5 &	-1.23 &	15.4 &	16.3 \\
\tableline
\multicolumn{15}{|c|}{ $\nu_s = 580$ Hz } \\
\tableline
	1.02 &	1.11 &	10.3 &	 0.7 &	 2.3 &	0.28 &	   2 &	 --- &	1.77 &	56.9 &	 6.6 &	11.0 &	-1.66 &	46.8 &	74.3 \\
	1.25 &	1.39 &	10.0 &	 0.7 &	 2.1 &	0.24 &	   3 &	10.1 &	2.01 &	77.7 &	29.1 &	12.8 &	-1.44 &	37.3 &	50.2 \\
	1.41 &	1.59 &	 9.8 &	 0.7 &	 1.8 &	0.22 &	   5 &	11.2 &	1.80 &	62.9 &	37.9 &	14.1 &	-1.32 &	31.8 &	39.3 \\
	1.46 &	1.66 &	 9.7 &	 0.7 &	 1.7 &	0.21 &	   6 &	11.7 &	1.73 &	58.2 &	38.6 &	14.6 &	-1.28 &	30.0 &	36.1 \\
	1.60 &	1.86 &	 9.2 &	 0.7 &	 1.2 &	0.18 &	  10 &	12.9 &	1.57 &	46.5 &	37.0 &	15.7 &	-1.20 &	25.6 &	29.0 \\
	1.67 &	1.95 &	 8.7 &	 0.7 &	 0.9 &	0.17 &	  14 &	13.5 &	1.50 &	39.4 &	34.0 &	16.1 &	-1.17 &	22.9 &	25.1 \\
\tableline
\multicolumn{15}{|c|}{ $\nu_s = 720$ Hz } \\
\tableline
	1.02 &	1.10 &	10.5 &	 0.7 &	 3.8 &	0.36 &	   2 &	 --- &	1.71 &	67.9 &	-12.0 &	11.6 &	-1.53 &	50.7 &	86.8 \\
	1.40 &	1.58 &	10.0 &	 0.7 &	 2.9 &	0.27 &	   5 &	11.0 &	1.86 &	85.5 &	37.6 &	14.6 &	-1.26 &	36.7 &	47.1 \\
	1.58 &	1.82 &	 9.4 &	 0.7 &	 2.1 &	0.24 &	   9 &	12.4 &	1.65 &	65.7 &	45.7 &	15.9 &	-1.16 &	30.7 &	36.0 \\
	1.60 &	1.86 &	 9.3 &	 0.7 &	 2.0 &	0.23 &	  10 &	12.6 &	1.63 &	63.0 &	45.6 &	16.1 &	-1.15 &	29.9 &	34.7 \\
	1.67 &	1.95 &	 8.8 &	 0.7 &	 1.4 &	0.21 &	  14 &	13.2 &	1.55 &	53.7 &	43.6 &	16.5 &	-1.13 &	27.1 &	30.3 \\

\tableline 
\end{tabular}
\end{table}

\clearpage

\begin{table}
\label{t:tableWS}
\caption{Precession frequencies, EOS WFF3}

\begin{tabular}{|rrrrrrr|rrrr|rrrr|}\tableline
\multicolumn{7}{|c|}{ } &
\multicolumn{4}{c|}{prograde} &
\multicolumn{4}{c|}{retrograde}\\
\tableline

$\bar{M}$	&$\bar{M}_B$	&$R$	&$I/{\bar{M}}$	&$\Phi_2$	&$j$	&$\zeta$ 
		&$r_{+}$	&$\nu_K$	&$\omega/2\pi$	&$\nu_p$ 
		&$r_{-}$	&$\nu_K$	&$\omega/2\pi$	&$\nu_p$ \\

\tableline
\multicolumn{15}{|c|}{ $\nu_s = 290$ Hz } \\
\tableline
	1.00 &	1.07 &	11.2 &	 0.8 &	 1.0 &	0.16 &	   1 &	 --- &	1.54 &	24.4 &	11.1 &	 --- &	-1.56 &	24.4 &	35.5 \\
	1.16 &	1.26 &	11.2 &	 0.8 &	 1.0 &	0.15 &	   2 &	 --- &	1.66 &	30.1 &	16.2 &	11.3 &	-1.67 &	29.2 &	40.1 \\
	1.35 &	1.50 &	11.1 &	 0.9 &	 1.0 &	0.13 &	   3 &	11.2 &	1.77 &	36.1 &	24.0 &	13.0 &	-1.46 &	23.5 &	29.2 \\
	1.41 &	1.57 &	11.0 &	 0.9 &	 0.9 &	0.13 &	   4 &	11.7 &	1.71 &	33.7 &	24.2 &	13.5 &	-1.41 &	22.2 &	26.7 \\
	1.62 &	1.84 &	10.7 &	 0.9 &	 0.7 &	0.12 &	   6 &	13.5 &	1.48 &	25.8 &	21.6 &	15.3 &	-1.24 &	17.7 &	19.9 \\
	1.85 &	2.16 &	 9.8 &	 0.8 &	 0.5 &	0.10 &	  13 &	15.5 &	1.28 &	18.1 &	16.5 &	17.2 &	-1.11 &	13.2 &	14.2 \\

\tableline
\multicolumn{15}{|c|}{ $\nu_s = 360$ Hz } \\
\tableline
	1.01 &	1.08 &	11.3 &	 0.8 &	 1.6 &	0.21 &	   1 &	 --- &	1.53 &	30.3 &	 9.3 &	 --- &	-1.56 &	30.3 &	47.1 \\
	1.12 &	1.22 &	11.2 &	 0.8 &	 1.5 &	0.19 &	   2 &	 --- &	1.62 &	35.5 &	15.1 &	11.3 &	-1.65 &	35.1 &	50.9 \\
	1.36 &	1.50 &	11.1 &	 0.9 &	 1.4 &	0.17 &	   3 &	11.2 &	1.80 &	46.5 &	26.6 &	13.3 &	-1.41 &	27.4 &	35.2 \\
	1.41 &	1.57 &	11.1 &	 0.9 &	 1.4 &	0.16 &	   3 &	11.6 &	1.73 &	43.4 &	27.3 &	13.8 &	-1.36 &	25.9 &	32.4 \\
	1.60 &	1.82 &	10.8 &	 0.9 &	 1.2 &	0.15 &	   6 &	13.2 &	1.52 &	34.3 &	26.6 &	15.4 &	-1.22 &	21.4 &	24.8 \\
	1.84 &	2.16 &	 9.9 &	 0.9 &	 0.7 &	0.12 &	  12 &	15.3 &	1.30 &	23.6 &	21.0 &	17.4 &	-1.09 &	16.0 &	17.3 \\

\tableline
\multicolumn{15}{|c|}{ $\nu_s = 580$ Hz } \\
\tableline
	1.03 &	1.10 &	11.7 &	 0.8 &	 4.2 &	0.34 &	   1 &	 --- &	1.47 &	47.8 &	-10.1 &	11.7 &	-1.51 &	47.5 &	87.6 \\
	1.22 &	1.33 &	11.5 &	 0.9 &	 4.1 &	0.30 &	   2 &	 --- &	1.63 &	62.6 &	 4.7 &	13.2 &	-1.37 &	41.1 &	63.8 \\
	1.43 &	1.60 &	11.3 &	 0.9 &	 3.7 &	0.27 &	   4 &	11.4 &	1.78 &	77.6 &	24.6 &	14.9 &	-1.23 &	34.6 &	47.1 \\
	1.62 &	1.84 &	11.0 &	 0.9 &	 3.2 &	0.24 &	   6 &	12.8 &	1.59 &	62.9 &	36.1 &	16.4 &	-1.13 &	29.6 &	36.7 \\
	1.70 &	1.95 &	10.8 &	 0.9 &	 2.8 &	0.23 &	   7 &	13.4 &	1.52 &	56.6 &	38.0 &	17.1 &	-1.09 &	27.5 &	32.8 \\
	1.85 &	2.16 &	10.1 &	 0.9 &	 1.9 &	0.20 &	  12 &	14.7 &	1.38 &	44.0 &	35.7 &	18.2 &	-1.03 &	23.3 &	26.1 \\

\tableline
\multicolumn{15}{|c|}{ $\nu_s = 720$ Hz } \\
\tableline
	1.01 &	1.08 &	12.2 &	 0.9 &	 7.1 &	0.45 &	   1 &	 --- &	1.37 &	54.5 &	-33.8 &	12.6 &	-1.36 &	49.2 &	99.9 \\
	1.40 &	1.55 &	11.6 &	 0.9 &	 6.2 &	0.35 &	   3 &	 --- &	1.73 &	93.4 &	 1.5 &	15.4 &	-1.17 &	39.6 &	57.9 \\
	1.46 &	1.63 &	11.5 &	 1.0 &	 5.9 &	0.34 &	   4 &	11.5 &	1.78 &	100.4 &	 8.7 &	15.9 &	-1.14 &	37.8 &	53.3 \\
	1.61 &	1.82 &	11.2 &	 1.0 &	 5.2 &	0.31 &	   6 &	12.5 &	1.65 &	86.6 &	32.5 &	17.0 &	-1.07 &	33.9 &	44.0 \\
	1.82 &	2.12 &	10.5 &	 0.9 &	 3.5 &	0.26 &	  11 &	14.1 &	1.46 &	64.8 &	45.0 &	18.6 &	-1.00 &	28.2 &	32.9 \\
	1.86 &	2.17 &	10.2 &	 0.9 &	 3.0 &	0.25 &	  12 &	14.4 &	1.43 &	60.3 &	44.7 &	18.8 &	-0.99 &	27.0 &	31.0 \\

\tableline 
\end{tabular}
\end{table}

\clearpage

\begin{table}
\label{t:tableC}
\caption{Precession frequencies, EOS C}

\begin{tabular}{|rrrrrrr|rrrr|rrrr|}\tableline
\multicolumn{7}{|c|}{ } &
\multicolumn{4}{c|}{prograde} &
\multicolumn{4}{c|}{retrograde}\\
\tableline

$\bar{M}$	&$\bar{M}_B$	&$R$	&$I/{\bar{M}}$	&$\Phi_2$	&$j$	&$\zeta$ 
		&$r_{+}$	&$\nu_K$	&$\omega/2\pi$	&$\nu_p$ 
		&$r_{-}$	&$\nu_K$	&$\omega/2\pi$	&$\nu_p$ \\

\tableline
\multicolumn{15}{|c|}{ $\nu_s = 290$ Hz } \\
\tableline
	1.00 &	1.07 &	12.7 &	 0.9 &	 1.6 &	0.19 &	   1 &	 --- &	1.28 &	20.1 &	 6.0 &	 --- &	-1.30 &	20.1 &	31.9 \\
	1.26 &	1.36 &	12.3 &	 1.0 &	 1.4 &	0.16 &	   2 &	 --- &	1.50 &	28.6 &	15.4 &	12.3 &	-1.52 &	28.2 &	38.7 \\
	1.40 &	1.53 &	12.0 &	 1.0 &	 1.3 &	0.15 &	   3 &	 --- &	1.63 &	34.0 &	21.2 &	13.5 &	-1.39 &	24.0 &	30.6 \\
	1.44 &	1.59 &	12.0 &	 1.0 &	 1.2 &	0.14 &	   3 &	12.0 &	1.67 &	35.9 &	23.0 &	13.9 &	-1.36 &	22.9 &	28.6 \\
	1.60 &	1.79 &	11.6 &	 1.0 &	 1.0 &	0.13 &	   5 &	13.3 &	1.50 &	28.8 &	22.3 &	15.3 &	-1.24 &	19.1 &	22.3 \\
	1.86 &	2.14 &	10.3 &	 0.9 &	 0.5 &	0.10 &	  12 &	15.6 &	1.28 &	18.5 &	16.7 &	17.3 &	-1.10 &	13.4 &	14.5 \\

\tableline
\multicolumn{15}{|c|}{ $\nu_s = 360$ Hz } \\
\tableline
	1.01 &	1.08 &	12.8 &	 1.0 &	 2.5 &	0.24 &	   1 &	 --- &	1.27 &	24.9 &	 3.0 &	 --- &	-1.30 &	24.9 &	42.6 \\
	1.25 &	1.36 &	12.4 &	 1.0 &	 2.2 &	0.20 &	   2 &	 --- &	1.48 &	34.9 &	14.1 &	12.7 &	-1.46 &	32.6 &	47.4 \\
	1.40 &	1.54 &	12.1 &	 1.0 &	 1.9 &	0.18 &	   3 &	 --- &	1.62 &	42.3 &	22.4 &	13.9 &	-1.34 &	27.8 &	36.7 \\
	1.46 &	1.61 &	12.0 &	 1.0 &	 1.9 &	0.17 &	   3 &	12.0 &	1.67 &	45.3 &	25.4 &	14.4 &	-1.30 &	26.2 &	33.6 \\
	1.59 &	1.77 &	11.7 &	 1.0 &	 1.6 &	0.16 &	   5 &	13.0 &	1.54 &	38.4 &	26.6 &	15.4 &	-1.22 &	23.0 &	27.8 \\
	1.86 &	2.14 &	10.3 &	 0.9 &	 0.8 &	0.12 &	  12 &	15.4 &	1.30 &	23.8 &	21.1 &	17.6 &	-1.08 &	16.1 &	17.5 \\

\tableline
\multicolumn{15}{|c|}{ $\nu_s = 580$ Hz } \\
\tableline
	1.03 &	1.10 &	13.6 &	 1.0 &	 7.4 &	0.41 &	   1 &	 --- &	1.18 &	37.8 &	-20.7 &	 --- &	-1.22 &	37.8 &	79.1 \\
	1.14 &	1.22 &	13.3 &	 1.0 &	 6.9 &	0.38 &	   1 &	 --- &	1.29 &	45.0 &	-13.9 &	13.4 &	-1.31 &	43.7 &	82.4 \\
	1.40 &	1.53 &	12.6 &	 1.0 &	 5.5 &	0.31 &	   3 &	 --- &	1.53 &	64.5 &	 8.3 &	15.1 &	-1.20 &	37.0 &	55.1 \\
	1.54 &	1.70 &	12.2 &	 1.0 &	 4.7 &	0.28 &	   4 &	12.2 &	1.67 &	76.5 &	22.6 &	16.1 &	-1.14 &	33.1 &	44.9 \\
	1.79 &	2.03 &	11.2 &	 1.0 &	 3.0 &	0.23 &	   8 &	14.1 &	1.45 &	54.1 &	37.4 &	17.9 &	-1.04 &	26.2 &	31.0 \\
	1.86 &	2.14 &	10.6 &	 0.9 &	 2.1 &	0.20 &	  12 &	14.8 &	1.37 &	45.5 &	36.1 &	18.4 &	-1.02 &	23.6 &	26.7 \\

\tableline
\multicolumn{15}{|c|}{ $\nu_s = 720$ Hz } \\
\tableline
	1.09 &	1.16 &	14.5 &	 1.1 &	12.9 &	0.53 &	   1 &	 --- &	1.10 &	44.8 &	-42.6 &	14.7 &	-1.14 &	43.6 &	98.1 \\
	1.41 &	1.53 &	13.0 &	 1.1 &	 9.3 &	0.40 &	   3 &	 --- &	1.46 &	76.8 &	-14.3 &	16.1 &	-1.10 &	39.8 &	63.2 \\
	1.60 &	1.78 &	12.3 &	 1.1 &	 7.3 &	0.34 &	   5 &	12.5 &	1.65 &	96.1 &	14.0 &	17.4 &	-1.04 &	35.2 &	48.4 \\
	1.87 &	2.15 &	10.8 &	 0.9 &	 3.6 &	0.26 &	  11 &	14.4 &	1.42 &	63.1 &	44.7 &	19.0 &	-0.97 &	27.5 &	31.9 \\

\tableline 
\end{tabular}
\end{table}

\clearpage

\begin{table}
\label{t:tableL}
\caption{Precession frequencies, EOS L}

\begin{tabular}{|rrrrrrr|rrrr|rrrr|}\tableline
\multicolumn{7}{|c|}{ } &
\multicolumn{4}{c|}{prograde} &
\multicolumn{4}{c|}{retrograde}\\
\tableline

$\bar{M}$	&$\bar{M}_B$	&$R$	&$I/{\bar{M}}$	&$\Phi_2$	&$j$	&$\zeta$ 
		&$r_{+}$	&$\nu_K$	&$\omega/2\pi$	&$\nu_p$ 
		&$r_{-}$	&$\nu_K$	&$\omega/2\pi$	&$\nu_p$ \\

\tableline
\multicolumn{15}{|c|}{ $\nu_s = 290$ Hz } \\
\tableline
	1.38 &	1.49 &	15.3 &	 1.6 &	 5.2 &	0.23 &	   1 &	 --- &	1.14 &	26.3 &	 4.6 &	 --- &	-1.17 &	26.3 &	43.4 \\
	1.49 &	1.61 &	15.3 &	 1.6 &	 5.3 &	0.22 &	   1 &	 --- &	1.18 &	29.0 &	 7.4 &	15.4 &	-1.20 &	28.6 &	45.1 \\
	1.88 &	2.10 &	15.4 &	 1.8 &	 5.3 &	0.19 &	   3 &	15.4 &	1.31 &	39.7 &	18.6 &	18.8 &	-0.99 &	21.7 &	29.0 \\
	2.19 &	2.49 &	15.3 &	 1.9 &	 4.9 &	0.18 &	   5 &	17.8 &	1.13 &	31.3 &	21.1 &	21.5 &	-0.87 &	17.8 &	21.7 \\
	2.71 &	3.22 &	14.2 &	 1.8 &	 3.0 &	0.14 &	  13 &	22.2 &	0.90 &	19.6 &	17.2 &	25.9 &	-0.73 &	12.4 &	13.6 \\

\tableline
\multicolumn{15}{|c|}{ $\nu_s = 360$ Hz } \\
\tableline
	1.29 &	1.39 &	15.5 &	 1.5 &	 8.0 &	0.31 &	   1 &	 --- &	1.09 &	29.4 &	-4.7 &	 --- &	-1.12 &	29.4 &	55.2 \\
	1.45 &	1.57 &	15.5 &	 1.6 &	 8.3 &	0.29 &	   1 &	 --- &	1.14 &	34.5 &	 0.1 &	15.8 &	-1.14 &	32.3 &	55.5 \\
	1.94 &	2.16 &	15.5 &	 1.8 &	 8.2 &	0.24 &	   3 &	15.6 &	1.29 &	49.8 &	17.5 &	19.9 &	-0.93 &	24.0 &	32.9 \\
	2.01 &	2.25 &	15.5 &	 1.8 &	 8.1 &	0.24 &	   4 &	16.2 &	1.25 &	47.3 &	19.8 &	20.5 &	-0.90 &	23.0 &	30.8 \\
	2.34 &	2.69 &	15.3 &	 1.9 &	 7.1 &	0.21 &	   6 &	18.7 &	1.08 &	36.9 &	24.3 &	23.3 &	-0.80 &	18.9 &	22.9 \\
	2.71 &	3.22 &	14.3 &	 1.9 &	 4.8 &	0.18 &	  13 &	21.8 &	0.93 &	26.1 &	21.9 &	26.4 &	-0.71 &	14.7 &	16.4 \\

\tableline
\multicolumn{15}{|c|}{ $\nu_s = 580$ Hz } \\
\tableline
	1.41 &	1.51 &	16.8 &	 1.8 &	24.7 &	0.52 &	   1 &	 --- &	1.01 &	47.6 &	-46.5 &	18.5 &	-0.91 &	35.3 &	75.4 \\
	2.07 &	2.32 &	16.2 &	 2.0 &	23.1 &	0.40 &	   4 &	16.3 &	1.26 &	84.3 &	-10.5 &	23.3 &	-0.77 &	28.2 &	41.3 \\
	2.22 &	2.51 &	16.0 &	 2.0 &	21.9 &	0.37 &	   5 &	17.1 &	1.21 &	78.3 &	 9.0 &	24.5 &	-0.73 &	26.5 &	36.6 \\
	2.40 &	2.75 &	15.8 &	 2.0 &	19.9 &	0.35 &	   7 &	18.2 &	1.14 &	70.4 &	24.5 &	25.9 &	-0.70 &	24.4 &	31.7 \\
	2.72 &	3.21 &	14.9 &	 2.0 &	13.9 &	0.30 &	  12 &	20.6 &	1.01 &	53.5 &	35.8 &	28.2 &	-0.65 &	20.5 &	24.1 \\

\tableline
\multicolumn{15}{|c|}{ $\nu_s = 720$ Hz } \\
\tableline
	1.55 &	1.67 &	18.9 &	 2.1 &	47.4 &	0.68 &	   2 &	 --- &	0.90 &	54.5 &	-76.7 &	22.3 &	-0.73 &	32.3 &	69.6 \\
	2.00 &	2.21 &	17.3 &	 2.1 &	42.2 &	0.54 &	   3 &	 --- &	1.15 &	93.3 &	-69.3 &	24.8 &	-0.69 &	30.4 &	49.4 \\
	2.17 &	2.43 &	16.9 &	 2.1 &	39.4 &	0.51 &	   4 &	17.0 &	1.22 &	106.7 &	-51.8 &	25.9 &	-0.67 &	28.9 &	43.4 \\
	2.40 &	2.74 &	16.4 &	 2.1 &	34.8 &	0.46 &	   6 &	18.0 &	1.17 &	98.6 &	-1.8 &	27.5 &	-0.64 &	26.7 &	36.6 \\
	2.73 &	3.20 &	15.4 &	 2.1 &	24.6 &	0.39 &	  11 &	19.8 &	1.06 &	78.8 &	38.4 &	29.7 &	-0.61 &	23.2 &	28.4 \\

\tableline 
\end{tabular}
\end{table}

\begin{deluxetable}{rrrrrrrrr}
\tablecolumns{9}
\tablewidth{0pc}
\tablecaption{Best fit neutron star models for 4U 1728-34}
\tablehead{
\colhead{EOS}    &  \colhead{$\bar{M}$} &   \colhead{$\nu_s$}   
& \colhead{$\nu_K$} & \colhead{$\nu_p$}
& \colhead{$\nu_K$} & \colhead{$\nu_p$}
& \colhead{$\nu_K$} & \colhead{$\nu_p$}}
\startdata
WFF3 & 1.65 & 360 & 0.90 & 10.2 & 0.99 & 12.3 & 1.10 & 15.0 \\
     & 1.78 & 360 & 0.90 & 10.4 & 0.99 & 12.5 & 1.10 & 15.3 \\
     & 1.85 & 360 & 0.90 & 10.1 & 0.99 & 12.2 & 1.10 & 15.0 \\

C & 1.68 & 360 & 0.90 & 10.6 & 0.99 & 12.7 & 1.10 & 15.5 \\
  & 1.80 & 360 & 0.90 & 10.8 & 0.99 & 13.0 & 1.10 & 15.9 \\
  & 1.86 & 360 & 0.90 & 10.5 & 0.99 & 12.6 & 1.10 & 15.5 \\

L & 1.94 & 360 & 0.90 & 12.4 & 0.99 & 14.0 & 1.10 & 15.7 \\
  & 2.03 & 360 & 0.90 & 13.7 & 0.99 & 15.7 & 1.10 & 18.0 \\
  & 2.11 & 360 & 0.90 & 14.9 & 0.99 & 17.2 & 1.10 & 20.0 \\
\enddata
\end{deluxetable}

\begin{deluxetable}{rrrrr}
\tablecolumns{5}
\tablewidth{0pc}
\tablecaption{Best fit neutron star models for KS 1731-260}
\tablehead{
\colhead{EOS}    &  \colhead{$\bar{M}$} &   \colhead{$\nu_s$}   
& \colhead{$\nu_K$} & \colhead{$\nu_p$} }
\startdata
WFF3 & 1.61 & 260 & 1.20 & 13.2  \\
     & 1.80 & 260 & 1.20 & 13.4  \\

C & 1.65 & 260 & 1.20 & 13.7 \\
  & 1.80 & 260 & 1.20 & 14.0 \\
  & 1.86 & 260 & 1.20 & 13.4 \\

L & 1.70 & 260 & 1.20 & 13.6 \\

\vspace{3mm} \\

WFF3 & 1.72 & 520 & 1.20 & 23.8  \\
     & 1.80 & 520 & 1.20 & 25.0  \\

C & 1.77 & 520 & 1.20 & 25.1 \\
  & 1.86 & 520 & 1.20 & 25.5 \\

\enddata
\end{deluxetable}

\end{document}